# Spatio-temporal Granger causality: a new framework

Qiang Luo[1,2†],Wenlian Lu[3,4,5†], Wei Cheng[3,†], Pedro A. Valdes-Sosa[6], Xiaotong Wen[7], Mingzhou Ding[7] and Jianfeng Feng[2,3,4,5*]

[1]Department of Management, School of Information Systems and Management, National University of Defense Technology, Hunan 410073, P.R. China

[2]Shanghai Center for Mathematical Sciences, Fudan University, Shanghai 200433, P.R. China

[3]Centre for Computational Systems Biology, School of Mathematical Sciences, Fudan University, Shanghai 200433, P.R. China

[4]Centre for Scientific Computing, University of Warwick, Coventry CV4 7AL, United Kingdom

[5]Fudan University-Jinling Hospital Computational Translational Medicine Centre, Fudan University, Shanghai 200433, P.R. China

[6]Cuban Neuroscience Center, Ave 25 #15202 esquina 158, Cubanacan, Playa, Cuba

[7]J. Crayton Pruitt Family Department of Biomedical Engineering, University of Florida, Gainesville, Florida, US

[†] These authors contributed to this work equally.

[*]Correspondence should be addressed to Jianfeng Feng, Fudan University, Shanghai 200433, P.R. China. E-mail: jianfeng64@gmail.com.

## Abstract

That physiological oscillations of various frequencies are present in fMRI signals is the rule, not the exception. Herein, we propose a novel theoretical framework, spatio-temporal Granger causality, which allows us to more reliably and precisely estimate the Granger causality from experimental datasets possessing time-varying properties caused by physiological oscillations. Within this framework, Granger causality is redefined as a global index measuring the directed information flow between two time series with time-varying properties. Both theoretical analyses and numerical examples demonstrate that Granger causality is a monotonically increasing function of the temporal resolution used in the estimation. This is consistent with the general principle of coarse graining, which causes information loss by smoothing out very fine-scale details in time and space. Our results confirm that the Granger causality at the finer spatio-temporal scales considerably outperforms the traditional approach in terms of an improved consistency between two resting-state scans of the same subject. To optimally estimate the Granger causality, the proposed theoretical framework is implemented through a combination of several approaches, such as dividing the optimal time window and estimating the parameters at the fine temporal and spatial scales. Taken together, our approach provides a novel and robust framework for estimating the Granger causality from fMRI, EEG, and other related data.

## Introduction

Granger causality, a standard statistical tool for detecting the directional influence of system components, plays a key role in understanding systems behaviour in many different areas, including economics (Chen et al., 2011), climate studies (Evan et al., 2011), genetics (Zhu et al., 2010) and neuroscience (Ge et al., 2012; Ge et al., 2009; Guo et al., 2008; Luo et al., 2011). The concept of Granger causality was originally proposed by Wiener in 1956 (Wiener, 1956), and introduced into data analysis by Granger in 1969 (Granger, 1969). The idea can be briefly described as follows: If the historical information of time series A significantly improves the prediction accuracy of the future of time series B in a multivariate autoregressive (MVAR) model, then the Granger causality from time series A to B is identified. In classic Granger causality, time-invariant MVAR models are used to fit the experimental data of the observed time series.

However, a time-varying property is a common phenomenon in various systems. For example, the gene regulatory network in *Saccharomyces cerevisiae* was reported to evolve its topology (Luscombe et al., 2004) with respect to different stimuli or different life processes. A time-varying protein-protein interaction network for p53 was reported in (Tuncbag et al., 2009), and the authors subsequently suggested the use of a 4D view of a protein-protein interaction network, with time being the 4$^{th}$ dimension. In the primary visual cortex of anaesthetized macaque monkeys, ensembles of neurons have dynamically reorganized their effective connectivity moment to moment (Ohiorhenuan et al., 2010). The importance of a slow oscillation, such as the theta rhythm, in a neuronal system was analysed in (Smerieri et al., 2010). It should be pointed out that even if the time series data are observed to be weakly stationary (i.e., stationary in the second moment), the system configuration may be time-varying. A typical example of this is $X_t = a\cos(\omega t + U_t) + \xi_t$, where $t$ is time, $a$ and $\omega$ are constants, $U_t \sim U[-\pi,\pi]$ is a uniform distribution, and $\xi_t$ is noise. It is thus natural to consider time-varying systems and attempt to understand their impact on the estimation of Granger causality.

Analysing systems with time-varying structures has recently attracted greater interest, and many statistical methods have been proposed. An adaptive multivariate autoregressive model using short sliding time windows was proposed in (Ding et al., 2000) to deal with a non-stationary, event-related potential (ERP) time series. Inspecting the directed interdependencies of electroencephalography (EEG) data, a short time window approach to define time-dependent Granger causality was proposed in (Hesse et al., 2003). Time-varying Granger causality was also modelled using Markov-switching models in (Psaradakis et al., 2005). In these models, time-varying Granger causality was modelled using a hidden discrete Markov process with a finite state space. Wavelet-based time-varying Granger causality to establish the functional connectivity maps from fMRI data was suggested in (Sato et al., 2006). Considering the time-series data as independent and identically distributed observations, a method to infer the time-varying biological and social networks was proposed in (Ahmed and Xing, 2009), but this method did not provide the directional information of the time-varying relationship between variables. In (Havlicek et al., 2010; Sommerlade et al., 2012), the dual Kalman filter was used to establish time-varying Granger causality between non-stationary time series. These approaches extended the classic Granger causality analysis to a non-stationary case through adaptive multivariate autoregressive modelling under the assumption that the coefficients in the time-varying MVAR model can be modelled by a random walk. As a response to research dealing with the time-varying properties in the MVAR model, and the definition of Granger causality as a function with respect to time, we propose the use of a robust *global* index for measuring the direct information flow between time series, despite the time-varying properties. Granger causality is currently a popular model for this purpose, but classic Granger causality does not consider the time-varying properties of the data. Moreover, it is a widely held misconception that the longer the time series we have, the more reliable the results that are obtainable for Granger causality.

The aims of this paper are twofold:

(1) We answer the following question: What is the impact of the temporal scale in MVAR models on the resulting directional influence of Granger causality? For Gaussian variables, Granger causality is equivalent to the directed information transfer between variables. The question therefore becomes how the temporal scale in the MVAR model influences the estimation of the information flows between each variable within a system. In (Smith et al., 2011), the authors compared the performances of Granger causality analyses with different time lengths, and found that the longer the time series was, the better the performance. In their simulations, however, the underlying circuit stayed the same. In this paper, we investigate the effects of time-varying underlying circuits on a Granger causality analysis both mathematically and empirically.

(2) The second aim of this paper is to provide an efficient algorithm for estimating the global Granger causality index between two time series without any prior knowledge of the TV-MVAR model. It should be emphasised that there is a trade-off between the fineness of the change-point set and the accuracy of the estimation of the coefficients at each time window. Time windows that are too short might prevent a reliable estimation of the parameters. Time windows that are too long, on the other hand, might increase the probability of an incorrect inference of Granger causality. Based on Bayesian information criterion (BIC) and a change-point searching algorithm, we propose a method for determining the optimal size of a change-point set and the optimal change-points as a means to achieve the optimal balance between the fineness of the Granger causality and the accuracy of the model estimation. The theoretical results and algorithms were verified by estimating the average and cumulative Granger causalities on the simulated and experimental data, both of which confirmed that a finer change-point set provides a larger overall causality measurement.

To achieve the above goals, the effect of a time-varying causal structure on a Granger causality analysis was investigated mathematically, where the following notations were used. Consider two time series $x$ and $y$ over time window $[0,T]$. The *change-point set* $S_1 = \{0 = t_0 < t_1 < \cdots < t_m = T\}$ defines the time-varying property of the MVAR model as follows: at each time window $[t_{k-1}, t_k)$, the MVAR model is static, i.e., the interacting coefficients between variables are constants; in different time windows, however, these may differ. In this case, it becomes a time-varying MVAR (TV-MVAR) model. There are two alternatives for estimating the Granger causality from $y$ to $x$ in the TV-MVAR model with respect to the change-point set, $S_1$. One is to estimate the local Granger causality at each time window $[t_{k-1}, t_k)$ and then average them, which is called the *average Granger causality*, $F_{y\to x}^{(a,S_1)}$. The other is to average the variances of the residual errors locally at each small time window so that the *cumulative Granger causality*, $F_{y\to x}^{(c,S_1)}$, can be established by comparing the estimated variances of the residual errors of $x$ by considering whether $y$ can predict the future of $x$. The TV-MVAR model depends on the change-point set that divides the whole duration into finer time windows, as shown in Figure 1. We therefore need to address the relationship between the causality definition and the fineness of the change-point set in the TV-MVAR model.

We proved that both cumulative and average Granger causalities are generally monotonically increasing functions with respect to the fineness of the change-point set (see Figure 1 for a summary, and Appendices A, B, and C for theory proofs). That is, the finer the TV-MVAR model is, the larger the change-point set is, and the larger the (average and cumulative) Granger causalities that can be estimated. In particular, as shown in Theorems B1 and B4, under certain assumptions, the estimation of the coefficients in the coarser MVAR model is the (weighted) average among those of the finer model. Hence, if the "true" time-varying coefficients are nonzero but fluctuate at around zero, the "averaging" estimation may reduce the estimated Granger causality to zero and give an incorrect inference of Granger causality.

Empirically, we demonstrated the robustness of the proposed spatio-temporal Granger causality analysis by computing the Pearson's correlation coefficients between the Granger causality patterns using two scanning sessions on the same subject from the enhanced Nathan Kline Institute-Rockland Sample (see Materials and Methods). By considering the spatio-temporal details of the fMRI data for the TV-MVAR model, Granger causality has much greater consistency across two scanning sessions for the ***same*** subject. In particular, the correlation coefficient greatly increases from 0.3588 using classic Granger causality with a static MVAR model and region-wise estimation, to 0.6059 through our approach, which includes the optimal TV-MVAR model and voxel-wise estimation.

The theoretical results have also been confirmed using two experimental fMRI datasets: a resting-state dataset and a task-associated dataset. For the resting-state fMRI dataset, the classic Granger causality analysis failed to identify any significant causal connectivity to the precuneus. In comparison, at a finer-scale for the TV-MVAR model, our Granger causality approaches indicate that the precuneus serves as a hub for information transfer in the brain. Information flows between the precuneus and visual regions were revealed, which is consistent with an experimental setting in which the data were collected when the subjects' eyes were open. For the task-associated fMRI dataset, the estimation of the average Granger causality for the attention blocks was found to be significantly larger than that estimated through classic Granger causality based on a static MVAR model for the whole time series for all twelve subjects used in the experiment.

## Materials and Methods

*Generation of Time Series with Time-varying Causal Structure*

1) Generation of Time Series with Continuous Time-varying Causal Structure

Consider two time series and the effective interdependencies between them, as described using the TV-MVAR model with a constant noise level. The time series were generated through the following toy model:

$$\begin{pmatrix} x^{t+1} \\ y^{t+1} \end{pmatrix} = \begin{pmatrix} A_{11}(t) & A_{12}(t) \\ A_{21}(t) & A_{22}(t) \end{pmatrix} \begin{pmatrix} x^{t} \\ y^{t} \end{pmatrix} + \begin{pmatrix} n_{xy}^{t} \\ n_{yx}^{t} \end{pmatrix}, \tag{1}$$

where

$$A_{11}(t) = 0.1, A_{12}(t) = 0.5\left(\frac{t}{600} - 1\right) \cdot u_1,$$

$$A_{21}(t) = 0.5\left(1 - \frac{t}{400}\right) \cdot u_2, A_{22}(t) = 0.1\sqrt{2}.$$

We generated this toy model 100 times by randomly setting the parameters $u_1$ and $u_2$ according to a uniform distribution at an interval of [0,1]. For each model, the time series observations were generated for 1200 time steps. The parameters $A_{12}$ and $A_{21}$ correspond to the causal influences in the *Y*→*X* and *X*→*Y* directions, respectively. A significant nonzero causal coefficient indicates the causal influence in the corresponding direction. In this simulation, we specified a change in the causal coefficient from positive to negative.

2) Generation of Time Series with Stepwise Time-varying Causal Structure

Consider a TV-MVAR model of two components with only one directional causal influence, *X*→*Y*; namely, setting the corresponding coefficient $A_{21}$ to have nonzero values. This model was derived from Eq. (1) with the step-wise coefficients as follows:

$$A_{11}(t) = 0.1, A_{22}(t) = 0.1\sqrt{2},$$

$$A_{12}(t) = 0, A_{21}(t) = \begin{cases} 0.5u_1, & 0 < t \le t_1 \\ 0, & t_1 < t \le t_2 \\ -0.5u_1, & t_2 < t \le t_3 \\ 0, & t_3 < t \le T \end{cases} \tag{2}$$

where $t_1 = 215$, $t_2 = 415$, and $t_3 = 715$. We generated two time series with 1200 time points and repeated this generation 100 times by randomly setting the parameter $u_1$ from a uniform distribution at an interval of [0.5,1.5]. In this simulation, the causal coefficient $A_{12}$ for the *Y*→*X* direction was set to zero, and thus there was no causal influence from *Y* to *X*, and the causal coefficient $A_{21}$ varied across different time windows.

3) Generation of BOLD Signal with Time-varying Effective Connection

Herein, we simulated the fMRI time series of two brain regions, *X* and *Y*, for 400 s. By introducing a time-varying causal structure, the simulation scheme for the fMRI data in (Schippers et al., 2011) was adopted. First, a neuronal interaction (local field potential, or LFP) was simulated using a bi-dimensional first-order TV-MVAR model with a time step of 10 ms:

$$\begin{pmatrix} x^{t+1} \\ y^{t+1} \end{pmatrix} = A \begin{pmatrix} x^{t} \\ y^{t} \end{pmatrix} + \begin{pmatrix} n_{xy}^{t} \\ n_{yx}^{t} \end{pmatrix}, \tag{3}$$

where

$$A_{11} = 0.9,\ A_{22} = 0.9,$$

$$A_{12} = 0,\ A_{21} = \begin{cases} 0.5, & 0<\mathrm{t} \le 18000 \\ -0.5. & 18000<\mathrm{t} \le 40000 \end{cases}$$

The model had an causal influence from *X* to *Y* of a predetermined time-varying strength, $A_{21}$, with no influence from *Y* to *X*.

Second, both signals were convolved with the default hemodynamic response models from the SPM5 toolbox, and Gaussian noises were added as physiological noise in the BOLD response. The HRF was specified through seven model parameters: delay of response relative to onset (in seconds), delay of undershoot relative to onset (in seconds), dispersion of response, dispersion of undershoot, ratio of response to undershoot, onset (in seconds), and length of kernel. To investigate the effect of hemodynamic response variability on the Granger causality analysis, we systematically varied the delay of response ranging from 0 to 5 s. To mimic the neuronal delay between the cause-region to the effect-region, time series *Y* was shifted by 50 ms against *X* before the convolution of the HRF (Deshpande et al., 2010; Schippers et al., 2011; Smith et al., 2012).

Third, BOLD signals were generated by down-sampling the convolved time series by 2 Hz as a high sampling rate, and 1 Hz as a low sampling rate (resembling an acquisition rate (TR) of an MR-scanner), and Gaussian noise was again added as acquisition noise. After each step, the signals were normalized to zero means and unit variances. The total amount of noise added was 20%.

*Experimental fMRI Datasets*

1) Multiband Imaging Test-Retest Pilot Dataset

This set of fMRI data comes from the enhanced Nathan Kline Institute-Rockland Sample. The whole dataset consists of resting-state fMRI recordings from two sessions for seventeen subjects (healthy, aged 19–57, thirteen males and four females).

The fMRI data were collected using 3 Tesla, and forty slices were acquired for 900 volumes. Multiband echo planar imaging approaches enable the acquisition of fMRI data with unprecedented sampling rates (TR = 0.645 s) for full-brain coverage through an acquisition of multiple slices simultaneously at the same time. For more detailed information about this data set, please see the website at http://fcon_1000.projects.nitrc.org/indi/pro/eNKI_RS_TRT/FrontPage.html.

Data pre-processing was performed using DPARSF software (Yan and Zang, 2010). The first fifty volumes were discarded to allow for scanner stabilisation. Since multiple slices were excited simultaneously, a simple slice time correction might not work well. Given its short effective TR, such a correction is probably less important, and is therefore omitted in our data pre-processing. After the realignment for head-motion correction, the standard Montreal Neurological Institute (MNI) template provided by SPM2 was used for spatial normalization with a re-sampling voxel size of $3 \times 3 \times 3$ $mm^3$. After smoothing (FWHM = 8 mm), the imaging data were temporally filtered (band pass, 0.01–0.08 Hz) to remove the effects of a very low-frequency drift and high-frequency noises (e.g., respiratory and cardiac rhythms). An automated anatomical labelling (AAL) atlas (Tzourio-Mazoyer et al., 2002) was used to parcellate the brain into ninety regions of interest (ROIs). To verify the principle of voxel-level Granger causality, the brain was also divided into 1024 ROIs with around 45 voxels each according to a high-resolution brain atlas provided by

(Zalesky et al., 2010).

2) Resting-State fMRI Dataset

The resting-state fMRI dataset is a subset of a large database, called the 1000 Functional Connectomes Project (Biswal et al., 2010), which is freely accessible at www.nitrc.org/projects/fcon_1000/. The dataset provided by Buckner's group at Cambridge, USA, was used for the present study. This dataset consists of 198 healthy subjects (75 males and 123 females, aged 18–30). The fMRI data (TR = 3 s) were collected using 3 Tesla, and 47 slices were acquired for 119 volumes. Further details about this dataset can be found at the website provided above.

The first five volumes were discarded to allow for scanner stabilization. DPARSF (Yan and Zang, 2010), which is based on SPM8, was used for pre-processing the fMRI data, including slice-timing correction, motion correction, co-registration, grey/white matter segmentation, and spatial normalization into a Montreal Neurological Institute (MNI) space, then and re-sampled to $3 \times 3 \times 3$ $mm^3$. The waveform of each voxel was detrended and passed through a band-pass filter of 0.01 to 0.08 Hz. The data were smoothed spatially (FWHM = 8 mm). As a result, time series data with 114 time points from ninety brain regions (AAL-atlas) for 198 subjects were achieved.

3) fMRI Dataset for Attention Task

The dataset of an fMRI time series for an attention-task experiment was provided by the Ding Group at the University of Florida, USA (Wen et al., 2012), which consisted of twelve subjects who successfully completed the task (eight females and four males, aged 20–28). This experiment adopted a mixed blocked/event-related design. There were twelve attention blocks and twelve passive-view blocks, along with some fixation intervals. In each attention block, the subjects performed a trial-by-trail cued visual spatial-attention task. The fMRI data were collected using 3 Tesla, and 33 slices were acquired for 180 volumes for each of the six runs with TR 2s. The dataset was pre-processed by slice timing, motion correction, co-registration to an individual anatomical image, and normalization to the Montreal Neurological Institute (MNI) template, and then resampled to $3\times3\times3$ mm$^3$, using DPARSF. The hemodynamic response function (HRF) was convolved by the blocked rectangular function corresponding to the given experimental condition during the GLM analysis. For more detailed information about this dataset, please see (Wen et al., 2012).

For each attention block, there were thirty data points, lasting for 60 s. The task average response was removed from each attention block by subtracting the mean of the time series data across twelve attention blocks. The first five data points (10 s) were discarded to eliminate the transient effects. The temporal mean was removed for each attention block to meet the zero mean requirement of the Granger causality analysis. Therefore, we had 300 data points for the twelve attention blocks. Herein, the causality between the right intra-parietal sulcus (rIPS) and right temporal-parietal junction (rTPJ) was studied. Time series of nineteen and seventeen voxels were used for rIPS and rTPJ, respectively (Wen et al., 2012).

*Granger Causality in TV-MVAR*

For two time series $x^t$ and $y^t$, with $t = 1,2,\cdots,T$, define a change-point set as an increasing integer series of $1 = t_0 < t_1 < \cdots < t_{m-1} < t_m = T+1$, denoted by $S_1$. Consider the following piece-wise constant linear system to describe the directional influence from $y^t$ to $x^t$:

$$x^{t+1} = \bar{a}_1^{S_1}(k)x^t + \bar{b}_1^{S_1}(k)y^t + n(t), t_{k-1} \le t < t_k, k = 1,\cdots,m \tag{4}$$

where $\bar{a}_1^{S_1}(k)$ and $\bar{b}_1^{S_1}(k)$ are the estimated time-varying coefficients from $S_1$. In addition, when ignoring the directed causality from $y^t$ to $x^t$, Eq. (4) becomes

$$x^{t+1} = \tilde{a}_1^{S_1}(k)x^t + \tilde{n}(t), t_{k-1} \le t < t_k, k = 1,\cdots,m \tag{5}$$

where $\tilde{a}_1^{S_1}(k)$ is the estimated time-varying coefficient in this model. At the $k^{\text{th}}$ time window, the Granger causality can be defined locally as

$$F_{y\to x}^{(k,S_1)} = \log\left[\frac{\sum_{t=t_{k-1}}^{t_k-1}\mathrm{var}(\tilde{n}(t))}{\sum_{t=t_{k-1}}^{t_k-1}\mathrm{var}(n(t))}\right].$$

The average Granger causality with respect to $S_1$ can be estimated through the average of the Granger causalities at the time windows and weighted by the corresponding window lengths:

$$F_{y\to x}^{(a,S_1)} = \frac{1}{T}\sum_{k=1}^{m} F_{y\to x}^{(k,S_1)}(t_k - t_{k-1}). \tag{6}$$

If the length of each time window is uniform, it becomes

$$F_{y\to x}^{(a,S_1)} = \frac{1}{m}\sum_{k=1}^{m} F_{y\to x}^{(k,S_1)}.$$

An alternative way to compute Granger causality is cumulating the residual square errors across all time windows. This is called *cumulative Granger causality* with respect to $S_1$, and can be estimated by

$$F_{Y\to X}^{(c,S_1)} = \log\left[\frac{\sum_{t=1}^{T}\mathrm{var}(\tilde{n}(t))}{\sum_{t=1}^{T}\mathrm{var}(n(t))}\right] = \log\left[\frac{\sum_{k=1}^{m}\sum_{t=t_{k-1}}^{t_k-1}\mathrm{var}(\tilde{n}(t))}{\sum_{k=1}^{m}\sum_{t=t_{k-1}}^{t_k-1}\mathrm{var}(n(t))}\right]. \tag{7}$$

In particular, if random variable $y^t$ is stochastically orthogonal to $x^t$ at each time, i.e., $E\left[\left(x^t - Ex^t\right)\left(y^t - Ey^t\right)\right] = 0$ for all *t*, the cumulative Granger causality can be estimated as

$$F_{Y\to X}^{(c,S_1)} =$$

$$\log\left[\frac{\sum_{k=1}^{m}\sum_{t=t_{k-1}}^{t_k-1}[a_1(t)-\tilde{a}_1^{S_1}(k)]^2\,\mathrm{var}(x^t)+\sum_{k=1}^{m}\sum_{t=t_{k-1}}^{t_k-1}[b_1(t)]^2\,\mathrm{var}(y^t)+\sum_{k=1}^{m}\sum_{t=t_{k-1}}^{t_k-1}\mathrm{var}(n(t))}{\sum_{k=1}^{m}\sum_{t=t_{k-1}}^{t_k-1}[a_1(t)-\overline{a}_1^{S_1}(k)]^2\,\mathrm{var}(x^t)+\sum_{k=1}^{m}\sum_{t=t_{k-1}}^{t_k-1}[b_1(t)-\overline{b}_1^{S_1}(k)]^2\,\mathrm{var}(y^t)+\sum_{k=1}^{m}\sum_{t=t_{k-1}}^{t_k-1}\mathrm{var}(n(t))}\right].$$

For details on the derivative of the Granger causality expressions, please see Appendix A. Herein, only a first-order regression model with one-dimensional variables is considered, but the approach and resulting work on a general high-order and high dimensional TV-MVAR model will be discussed in a future paper.

Since $F_{y\to x}^{(k,S_1)}$ obeys an $F$-distribution after proper scaling in each time window, the average Granger causality defined above can be considered in the null hypothesis as the summation of $m$ independent $F$-distributed random variables whose degrees-of-freedom can be given according to the number of free parameters and the length of each time window, particularly 1 and $t_k - t_{k-1} - 3$. Therefore, the $p$-value for the significance of average Granger causality can be calculated. Similarly, cumulative Granger causality as defined above also obeys an $F$-distribution with degrees-of-freedom of $m$ and $T$-2$m$-1.

*Optimal Time Window Division*

In practice, the ***true*** time-varying structure of the data is unknown. In particular, we do not know how many change-points there are, or the length of each time window. Therefore, an algorithm for time-window division is necessary. Equivalently, we are searching for the optimal change-point set. The optimal time-window division indicates a trade-off between the satisfactory accuracy of the model parameter estimation and the lossless causal information established by the model. Mathematically, consider the following step-wise TV-MVAR model

$$X(t+1) = \sum_{k=1}^{m} a_1^k X(t) I_{[t_{k-1},t_k)} + n(t), \tag{8}$$

where $I_{[t_{k-1},t_k]}$ is the characteristic function of time window $[t_{k-1},t_k)$, $n(t)$ is a Gaussian white noise term, $a_1^k$ represents a (constant) coefficient in the $k^{\text{th}}$ interval, and

$$S(m) = \{t_1,\cdots,t_{m-1} \,|\, 1 = t_0 < t_1 < \cdots < t_{m-1} < t_m = T+1\}$$

is the change-point set. Given the change-points, the model can be fit into each time window as $\hat{a}_1^k$, and the variance of the residual errors can be estimated for each time window, denoted by $\hat{\Sigma}_k$. Therefore, the accuracy of the model can be defined based on the weighted average of the variances of the residuals in each time window as

follows:

$$err(S(m)) = \frac{1}{m}\sum_{k=1}^{m}(t_k - t_{k-1})\det(\hat{\Sigma}_k). \tag{9}$$

On the other hand, the information captured by this model can be measured based on the average Granger causality in all directions defined in the previous section, as noted by

$$agc(S(m)) = \frac{1}{2m}\sum_{k=1}^{m}(F_{y\to x}^{(k,S(m))} + F_{x\to y}^{(k,S(m))}). \tag{10}$$

To minimize the prediction error and maximize the detected causality information, the optimal window division can be derived by optimising the following cost function with the trade-off parameter $\lambda$

$$S_{opt}(m,\lambda) = \arg\min_{S(m)}\left(err(S(m)) + \lambda/agc(S(m))\right). \tag{11}$$

Given the trade-off parameter $\lambda_0$ and lower bound $l_0$ of the lengths of the divided time windows, the optimal change-points $S_{Opt}(m,\lambda_0)$ can be established by solving the following constrained optimization problem

$$\min_{S(m)}\left(err(S(m)) + \lambda_0/agc(S(m))\right) \tag{12}$$

$$\text{s.t. } t_k - t_{k-1} \geq l_0 \text{ for all } k = 1,2,\ldots,m.$$

A constrained condition is required for a reliable estimation of the model coefficients in Eq. (8) at each divided time window. This constrained optimization problem can be solved based on the optimization functions provided in Matlab. In this paper, we used the *fmincon* function for a nonlinear constrained optimization problem.

To determine the parameter, we search for the optimal change-point set $S_{opt}(m,\lambda)$ for different $\lambda \in [\lambda_1, \lambda_2]$, and then calculate the Bayesian information criterion (BIC) for this change-point set as follows:

$$BIC(m,\lambda) = -2\sum_{k=1}^{m} LLF_k + 2^2 m\log(T+1), \tag{13}$$

where $LLF_k$ stands for the log likelihood function established for the $k^{th}$ window. The first step is searching for the optimal change-point set with a series of given time windows, $m \in [0,1,2,\cdots,m_0]$, and trade-off parameter, $\lambda \in [\lambda_1, \lambda_2]$. The second step is to compare the BIC values established by different change-point sets generated from the first step, and the one with the smallest BIC is then selected to define the optimal time window. Therefore, using the fixed upper bound of the number of time windows, denoted by $m_0$, the algorithm for the optimal time window division can be described as follows:

*Algorithm for optimal time window division*

For $\lambda =$ from $\lambda_1$ to $\lambda_2$

  For $m =$ from 1 to $m_0$

    Establish $S_{opt}(m,\lambda)$ by solving the constrained optimization problem (12)

  End

  Calculate the BIC for each $S_{opt}(m,\lambda)$ by (13)

End

Find the optimal $S_{opt}(m_{opt},\lambda_{opt})$ with the smallest BIC

*Spatio-temporal Granger Causality*

Furthermore, both spatial and temporal fineness are taken into the MVAR model. The idea of a spatial finer-scale for Granger causality estimation is similar to that of the time-varying Granger causality mentioned above. Consider a dataset of fMRI BOLD signals from *m* voxels in ROI *A*, and *n* voxels in ROI *B*. For each pair of voxels in these two ROIs, the Granger causality between the voxel pair is calculated for each subject, denoted by $F_{ij}$, from the $i^{th}$ voxel in ROI *A* to the $j^{th}$ voxel in ROI *B*; the *global* Granger causality from ROI $A$ to ROI $B$, namely, *voxel-level Granger causality*, is then defined as follows:

$$F_{A\to B} = \frac{1}{mn}\sum_{i\in ROI_A}\sum_{j\in ROI_B} F_{ij}.$$

Furthermore, the temporal and the spatially fine-scales are combined together to give the optimal estimation of Granger causality by looking into the temporal details for each pair of voxels, which is called *spatio-temporal Granger causality* (stGC):

$$F_{A\to B}^{(e,S)} = \frac{1}{mn}\sum_{i\in ROI_A}\sum_{j\in ROI_B} F_{ij}^{(e,S)},$$

with $e = a$ or c for average and cumulative (time-varying) Granger causalities, respectively. In comparison, classic Granger causality usually estimates the causality between two ROIs by averaging the time series data among all voxels for each ROI with a static MVAR model.

A Matlab package for the estimation of the spatio-temporal GC is available at http://www.dcs.warwick.ac.uk/~feng/causality.html.

## Results

*Monotonicity of Granger Causality with Respect to Change-point Set*

To demonstrate the monotonicity of the proposed Granger causality measurements, the proposed algorithms were applied to a simulation dataset with a continuous time-varying causal structure. We used three different time-window lengths of 50, 200, and 400, and the corresponding change-point sets for these time windows denoted as $S_i, i = 1,2,3$ , respectively. Theorems B1 and B4 in Appendix B show that $F_{Y\to X}^{a,S_1} \geq F_{Y\to X}^{a,S_2} \geq F_{Y\to X}^{a,S_3}$ and $F_{Y\to X}^{c,S_1} \geq F_{Y\to X}^{c,S_2} \geq F_{Y\to X}^{c,S_3}$ hold if the parameters are precisely estimated since $S_1 \supset S_2 \supset S_3$ . To demonstrate this, 95% confidence intervals of $D_1^a = F_{Y\to X}^{a,S_1} - F_{Y\to X}^{a,S_2}$ , $D_2^a = F_{Y\to X}^{a,S_2} - F_{Y\to X}^{a,S_3}$ , and $D_3^a = F_{Y\to X}^{a,S_1} - F_{Y\to X}^{a,S_3}$ were established for the causality results in 100 runs of the simulated toy model. Similarly, $D_i^c, i = 1,2,3$ were defined, and their confidence intervals established. From Table 1, we can see that the estimated Granger causality for the same pair of time series decreases with respect to the length of the time windows. That is, the more change-points that are used in the TV-MVAR model, i.e., the finer the model is, the larger the Granger causality that can be estimated.

To show the accuracies of the model estimation, we compared the differences in the variances of the model residual errors given by different algorithms, including the static MVAR model by the whole time series, denoted by $Err_{[1,1200]}$, and the average variances of the model residual errors for the TV-MVAR model over the time windows, $\overline{Err_{S_i}}$, as $D_i = Err_{[1,1200]} - \overline{Err_{S_i}}$ for $i = 1,2,3$. As shown in Figure 2A, the TV-MVAR models with different time-window lengths all have smaller variances than the static MVAR model fit onto the whole time series for all 100 toy models, i.e., $D_i > 0$ holds for the 100 toy model runs. Among the models with different time window sets, the one with the smallest time-window length, which had $S_1$ as the change-point set, provided the most accurate estimation of the simulated time series.

*Significance of Granger Causality*

To compare the significance of the results detected by our time-varying Granger causality approach with those detected by classic Granger causality, we applied these algorithms on the simulation dataset with a stepwise causal structure. When the *p*-value was lower than the threshold, a significant directional influence was detected. In this simulation setup, a causal influence existed from *X* to *Y*, but not from *Y* to *X*. The usual definitions of the truth positive (TP), false positive (FP), truth negative (TN) and false negative (FN) were used. In addition, the maximum number of time windows was set to $m_0 = 5$, and the trade-off parameter ranged from $\lambda_1 = 0.02$ to $\lambda_2 = 1$ with a step size of 0.02. Five types of Granger causalities, classic Granger causality (classic GC), average Granger causality (average GC), cumulative Granger causality (cumulative GC), average Granger causality with optimally divided time windows (Opt average GC), and cumulative Granger causality with optimally divided time windows (Opt cumulative GC) were calculated based on the simulation data using a significance test.

As shown in Table 2, the classic GC failed to identify any causal influence between these two time series. Cumulative GC and average GC provided better results in terms of higher TP and TN rates than classic GC. Compared to other algorithms, average GC and cumulative GC with optimally time window division provided the best performances in terms of the TP and TN rates among all of the causalities. In particular, as shown in Theorems B1 and B4, under our assumption, in the coarser MVAR model, the estimation of the coefficients is the (weighted) average of those of the finer model. As an intuitive interpretation, if the "true" time-varying coefficients are nonzero but have fluctuating signs, for example, they equal 1 at the first time interval and -1 at the last time interval with same length, the "averaging" estimation becomes zero owing to the neutralisation, even if the coefficient parameters are precisely estimated. Thus, in the static MVAR model, we will incorrectly infer that no Granger causality exists. A similar argument holds for a comparison of the finer and coarser MVAR models. Therefore, the coarseness of the TV-MVAR model might increase the probability of an incorrect inference of Granger causality.

To demonstrate the performance of the optimal GC estimations using time windows with equal lengths, we compared the accuracies of the results given by the optimal GCs with different time-window lengths. We found that better performances were achieved if their time-window division was similar to the ***real*** structure of the simulation data. Since the real change-points of this simulation were 215, 415, and 715, both algorithms presented better results when the time-window lengths were 100 or 300. In comparison, the performances worsened when the time-window lengths were either longer or shorter.

To test whether the larger magnitude of Granger causality estimated by the optimal

GCs increases the false positive (FP) rate, Table 2 also lists the FP rates given by different GCs with different window lengths. We found that the FP rates of both cumulative GC and average GC with different window lengths were zero when the threshold of the significance was $10^{-12}$ (for the $F$ statistics). Therefore, the FP rates of these two algorithms did not increase with respect to the GC values, as the lengths of the time windows shortened. Since the degrees-of-freedom of the $F$ statistics depended on the number of change-points, the GC value increased with shorter time windows. However, since the corresponding $F$ distribution also changed with the number of change-points, the FP rates might not have increased.

To assess the rationality of the BIC-based optimal time-window dividing algorithm, the BIC values were also reported and compared among the simulations. As shown in Table 2, good BIC values were achieved when the change-point set for the average and cumulative GCs was similar to the true structure of the simulated data. This suggests that the BIC values can work for choosing change-points to achieve the best performance. We chose the change-point set optimally instead of using time windows with equal length. Table 2 also shows that, compared to algorithms with an equal time window division, algorithms with optimally change-point searching provide better TP rates, but slightly worse FP rates, namely, 3% for the opt cumulative GC, and 2% for the opt average GC, in the simulation data. Figure 2B shows that the real change-points for 100 simulations using the proposed BIC-based optimal algorithm were successfully identified for most of the simulations.

To compare the computational complexities among the different algorithms, we reported the running time of each algorithm on the simulated dataset. As listed in Table 3, because the method for optimally dividing the time windows is very time-consuming, the greater the number of time windows we used, the greater the amount of time that was required to run the algorithm. In practice, since the underlying time-varying structure of the data is unknown, we can either run the optimal time-window dividing algorithm, or try different time-window lengths and select the optimal length through a comparison of their BICs.

*Effect of Regional Variation in HRF on Granger Causality Analysis*

The effects of the HRF delay of the response on the Granger causality analysis were simulated by setting the delay of the response relative to the onset of the HRF for brain region *X* as the parameter for brain region *Y* plus a delay ranging from 0 to 3 s. Therefore, the underlying causal influence existed from *X* to *Y*, but the HRF of cause-region *X* was slower than that of effect-region *Y*. We refer to this delay as the *opposite HRF delay*. The longer this delay is, the more difficult it is for a Granger causality analysis to detect the causal influence correctly. Setting the threshold of the *p*-value for a significant causality as $10^{-6}$, Figure 2C plots the TP and FP rates of different algorithms as the opposite HRF delay varies from 0 to 3 s. Opt average GC and opt cumulative GC performed similarly during the simulation, and therefore, only the results for opt average GC are shown. We can see that the optimal Granger causality performed well as long as the opposite HRF delay was less than 100 ms. When the opposite HRF delay was greater than 100 ms, the FP rate increased and the TP rate dropped rapidly. The TP rate increased again when the opposite HRF delay exceeded 0.4 s because an opposite HRF delay was generated by changing the shape of the HRF (Deshpande et al., 2010). We also carried out our simulations by changing the onset time of the HRF instead of changing its shape, and obtained similar results (data not shown).

To test whether the opposite delay in the HRF can be corrected for the optimal GC algorithms, we realigned the simulated BOLD signal according to the HRF delay between two regions by assuming that the regional HRF delay, especially the relative HRF delay between these two regions, can be accurately estimated. For example, when the HRF of region *Y* was estimated to be 3 s faster than that of region *X*, we realigned the time series of region *X* against that of region *Y* by discarding the first three and last three data points of the time series of regions *X* and *Y*, respectively, at a sampling rate of 1 Hz. The regional HRF delay was simulated by setting different parameters of the response relative to the onset in the canonical HRF in the SPM with the default settings, and the opposite HRF delay was varied from 2 to 5 s. Figure 3A shows the results given by the GC algorithm with BOLD signal realignment, when the sampling rate of the BOLD signal was 1 Hz. Setting a threshold of $10^{-9}$ for the p-value, classical GC failed to detect any causality in this case, but the proposed GC algorithms achieved much better TP and FP rates. However, the BOLD realignment worked for those integer HRF delays matching the sampling rate, but not for those delays that are not the integer times of the sampling period, which herein is 2.5, 3.5, and 4.5 s. Therefore, we tried to increase the sampling rate to 2 Hz, and simulated the BOLD signal again. In Figure 3B, without the BOLD realignment, the proposed GC algorithms failed to reliably estimate the causality because both the TP and FP rates are high. The BOLD realignment improved the performances of the proposed GC algorithms with a 100% TP rate and lower than 20% FP rate, as shown in Figure 3C. We can barely see the results for classic GC in Figure 3, since classic GC failed to detect any significant causal causality in all cases.

As demonstrated above, the proposed optimal GC algorithms may detect the right direction, the reversed direction, or the bi-direction of the causal influence between two regions as significant. However, what if there is no causal influence between the two regions? To test whether the down sampling and HRF convolution introduce false causal connections between pairs of regions without any causal influence in neuronal activities, the proposed algorithms were applied to other simulation data by setting the causal coefficient, $A_{21}$, in model (3) to zero, the neuronal delay to 50 ms, and the opposite HRF delay to 3 s. Setting the threshold of the $p$-value for significant causality as $10^{-6}$, the false positive rates for opt average GC and opt cumulative GC were 0.24% and 0.12%, respectively.

*Performance Comparison of Simulated fMRI Dataset*

First, classic Granger causality (classic GC), optimal GC approaches (opt average GC and opt cumulative GC), and dual Kalman filter cumulative GC (Dkf cumulative GC), which is defined in Appendix D, were applied to the simulated fMRI dataset described in the Materials and Methods section for a performance comparison. All results were obtained by repeating the simulation 100 times. Neither the neuronal delay nor the negative HRF delay was included in this simulation, as none of the lag-based methods work well in this case (Smith et al., 2012); however, such a performance comparison is informative when the lag-based method is applicable (Friston et al., 2012; Wen et al., 2012). When coefficient matrix $A$ was time-invariant, all approaches could detect the causality correctly, as expected. When the interaction coefficients were time-varying, in particular, with positive and negative values alternatively in different time intervals, as defined in Eq. (3), the optimal GC approaches were much more powerful than both classic GC and dual Kalman filter cumulative GC. To obtain a more global view of the results, the threshold of the $p$-values was varied from 0.05 to 0.001. We calculated the TP and FP rates of these three approaches, i.e., opt average GC, opt cumulative GC, and Dkf cumulative GC, accordingly, by repeating the simulation 100 times. As shown in Figure 2D, the proposed optimal GC approaches outperformed dual Kalman filter cumulative GC.

*Increased Test-retest Reliability Obtained from Multiband Resting-State Dataset*

Herein, the reliability of Granger causality can be measured based on the correlation between the results inferred for two series of scans of the same subjects. Granger causality was estimated between all directional pairs of brain regions, and the Pearson's correlation coefficients were then calculated between these causality measurements for the two series scans. For each scan in the multiband test-retest pilot dataset, the Granger causality for each direction was averaged over seventeen subjects to provide the group Granger causality. The correlations of the group Granger causality between two series of scans demonstrate the reliability of Granger causality. Larger correlations might result in a higher reliability. As shown in Figure 4, the correlations in the group Granger causality between two series of scans increased monotonically with respect to the number of change-points. A significant correlation ($r = 0.4751$, $p < 0.001$) in the group Granger causality between two series of scans was observed when the Granger causality was calculated by employing nineteen time windows, while the correlation was around 0.3105 in the classical case.

To further demonstrate the effect of the spatial fine-scale details on the Granger causal inference, we compared the correlations established by voxel-level Granger causality with those by classic Granger causality in 100 randomly selected regions from 1024 ROIs by averaging the time series in the same ROI. By calculating the voxel-level spatial Granger causality instead of the classic Granger causality, the correlation increased from 0.3588 to 0.5125. Furthermore, the combined effects of the temporal and spatial fine-scale details were demonstrated on the test-retest reliability of the Granger causality for 100 regions randomly selected from 1024 ROIs. In Figure 5, by calculating the spatio-temporal Granger causalities (stGC), the correlation ($r = 0.6059$, $p < 0.001$) between the two scans was significantly improved to 0.6059.

*Validating the Results from the Resting-State fMRI Dataset*

1) Monotonicity and significance of Granger causality

In this example, the Granger causality was estimated using time windows with different lengths. In each time series, the first eighty time points were divided into two sets of time windows, including eight time windows with ten time points per window, i.e., change-point set $S_1 = \{1,10,20,40,50,60,70,80\}$, and two time windows with forty time points per window, i.e., change-point set $S_2 = \{1,40,80\}$. The cumulative and average Granger causalities with $S_1$ and $S_2$ were estimated for all directions between all pairs of brain regions for each subject. A 95% confidence interval of the differences $D^a_{j\to i} = F^{a,S_1}_{j\to i} - F^{a,S_2}_{j\to i}$ and $D^c_{j\to i} = F^{c,S_1}_{j\to i} - F^{c,S_2}_{j\to i}$ was established for each possible direction $\{i \to j \mid i,j = 1,2,\cdots,90 \text{ and } i \neq j\}$. For all directions, the lower bounds of these differences were still larger than 0, which is exactly consistent with our theoretical results, as shown in Figure 6B.

The results of the average and cumulative Granger causalities were well correlated, as shown in Figure 6A. Actually, if the data are generated by the TV-MVAR model, which is perfectly static in each time window, the cumulative Granger causality is larger than the average Granger causality (See Theorem C1 in Appendix C). Under the null hypothesis of non-causality, both Granger causalities approach zero as the size of the data becomes sufficiently large. Moreover, the average Granger causality converges to zero quicker than the cumulative Granger causality (Theorem C2 in Appendix C), *i.e.*, the *p*-value of the significance of the average Granger causality may be smaller, as was also shown from the simulation results in the previous section (Table 2) in which the average Granger causality performed better than the cumulative Granger causality in terms of detecting the non-causality. Therefore, in the following, the average Granger causality is calculated.

As discussed in Appendix B (Corollary B5), some causal connectivity may be missed if the Granger causality is estimated using the static MVAR model for the whole time series, owing to the correlation of the causality measurement and the time-varying causal coefficients. We studied individually the correlations between the causality measurements, the sum of the absolute values of the estimated causal coefficients, and the absolute value of the sum of the estimated causal coefficients across all time windows defined by the change-point set ($S_1$) in the TV-MVAR model. For 198 subjects, the absolute values of the median of this summation were plotted in Figure 7A against the median of the Granger causality for each direction. This correlation between the average Granger causality and the sum of the causal coefficients decreased for finer time windows, as compared with the classic Granger causality. In contrast, this correlation increased when the absolute value of the median of the sum of the causal coefficients was considered (Figure 7B). As shown in Eq. (A2) in Appendix A, summing the positive and negative causal coefficients in different time windows may lead to an elimination of both positive and negative causal influences. In other words, the classic Granger causality, or Granger causality with a coarser-scale, tends to give a null prediction when the sum of the causal coefficients is near zero; however, a zero sum may be given by significant non-zero coefficients with different signs in different time windows.

The results for some particular examples are given in Figure 8. The classic Granger causality using the whole time series data gave near-zero[1] causality measurements when the summations of the causal coefficients across all time windows were near zero for the directions 'Precuneus_R→Hippocampus_R' and 'Thalamus_R→Precuneus_L'. However, both the average Granger causality and the classic Granger causality detected significant causality for the other three directions, as shown in Figure 8, since the sum of the causal coefficients across all time windows was larger than zero.

2) Granger causality mapping from the precuneus

The approaches discussed above were used to identify the Grange causality mapping from the precuneus, which is believed to be the core of many cognitive behaviours and self-conscience, and has been called the 'mind's eye' (Cavanna and Trimble, 2006), to other brain regions. The proposed average Granger causality with the optimal time-window dividing algorithm (AGC-OTWDA) was used for the resting state dataset by setting the maximum number of time-windows to three. The significance of the causality influence was detected through a statistical test (see Materials and Methods). In contrast, an analysis was also carried out for each subject using the classic Granger causality.

The classic Granger causality based on the whole time series failed to detect any significant[2] causal connectivity from the precuneus, while the AGC-OTWDA

[1]The magnitude of the causality measurement is significantly larger than 0 if the lower bound of the 95% confidence interval of the causality in 198 subjects is greater than 0.0002 for the classical Granger causality, and 0.0726 for the average Granger causality.

[2]A significant causality was identified when its $p$-value was less than 0.05 in at least

identified directional neural circuits centred at the precuneus, as shown in Figure 9. Since this dataset was collected when the subjects' eyes were open, the information flows from the precuneus and visual recognition network of the brain regions, marked in green in Figure 9, were very significant.

To ascertain that the relative variation of the HRF is not a significant confounding factor for the results of the precuneus, the cross-correlation function between the BOLD signals of two regions in each causal connection was examined. The peaks of the cross-correlation function appeared to have zero lag in more than 90% of the subjects for most of the pairs, except for those between the right precuneus (PCUN.R), the right Precental gyrus (preCG.R), and the opercular part of the right inferior frontal gyrus (IFGoperc.R), which had only 68% peaks with zero lag. Therefore, the relative variation of the HRF was not a significant factor in the causality results between the precuneus and the visual recognition network.

*Validating Results on the Attention-Task fMRI Dataset*

For the attention task, we detected the causality between rIPS and rTPJ. Granger causalities were estimated for all possible pairs of voxels and averaged as the spatial Granger causality. Two methods were used to calculate the Granger causality. One was to concatenate the time series data in each attention block together into a long data series, and then compute the Granger causality. The other was to calculate the Granger causality for each attention block and average them, i.e., the average Granger causality defined above. For comparison, we applied these two methods to estimate the Granger causality of two directions, rIPS→rTPJ and rTPJ→rIPS, for twelve subjects.

As shown in Figure 10, the average Granger causality is clearly larger than the classic Granger causality. For both directions, the differences between the causalities

---

73% of the subjects.

established through the two different methods were calculated for all twelve subjects, and a paired two-sample t-test was conducted to examine the difference, i.e., the average Granger causality subtracting the classic Granger causality. The right-tailed t-test suggested that the differences in both directions are significantly larger than 0 with $p$-values equal to $6.6482 \times 10^{-6}$ for rIPS→rTPJ, and $9.0040 \times 10^{-5}$ for rTPJ→rIPS. These results are consistent with our theoretical analysis, i.e., the average Granger causality analysis across many shorter time series provided by multi-trails provides larger measurements than a single long-term series observation.

## Discussion

*Danger of Smoothing Out Causal Information in Long-term Recordings*

When we have long-term recordings of two time series observations, how can we reliably estimate the Granger causality between the time series? A naive and intuitive approach to estimate the Granger causality is to apply all recordings into the MVAR model. This approach is based on the widely-accepted statistical belief that the more data that are used, the closer the result will be to the true value. However, in this paper, our theoretical analysis and numerical examples demonstrate that this may not be the case in an fMRI data analysis. The reliability of the statistical inference depends not only on how many datasets there are, despite their importance, but also on how finely the model describes the data.

In this paper, we discussed the effects of the fine-scaled details in the MVAR model on the Granger causality for detecting the directional information flows between time series data and applied the results to the fMRI data analysis. This effect was mathematically analysed, and it was concluded that both the temporal and spatial characteristics of the MVAR model affect the reliability of the Granger causality estimation. A smaller change-point set implies a coarser model, and a larger one implies a finer model. As we proved, the Granger causalities in the coarser model (with fewer change-points), including both the cumulative and average causalities, are smaller than those in the finer model (with more change-points). As demonstrated by the numerical simulations, the classic Granger causality becomes the lower bound of the average and cumulative Granger causalities (Corollaries B2 and B5), while the causality established using the real change-point set provides the upper bound. Our results demonstrate that the Granger causality depends on the model configuration, and thus ‘the devil is in the details’.

The Granger causality was proved to be equivalent to the transfer information (entropy) between Gaussian processes. It has been widely argued that the definition of the information strongly depends on the modelling configuration for the physical system (Jaynes, 1985). As argued by (Lloyd, 1989), coarse-grained modelling (such as imperfectly determined network evolution) may lead to information loss. Hence, the calculation of Granger causality, or the transfer of information, definitely suffers from the modelling configuration issue.

*Trade-off between Preciseness of Estimation and Fineness of Modelling*

For a given data set, if we use too many change-points for the TV-MVAR model to have a sufficient number of data points at each time window, we may obtain an inaccurate estimation of the coefficients for the model. In other words, a larger change-point set implies a finer model (possibly a larger Granger causality), but this may become an obstacle for the precise estimation of the Granger causality. Therefore, an optimal change-point set should be a trade-off between the preciseness of the statistic estimation and the fineness of the modelling. In this paper, we propose a novel algorithm for detecting the optimal change-point set based on the Bayesian information criterion (BIC). As illustrated through a numerical simulation, this algorithm can correctly identify the change-points in the model and increase the reliability and significance for a Granger causality analysis. However, compared to the models with time windows of equal length, the optimal time-window dividing algorithm was shown to be time consuming. When we focus on the global index measuring the directed information flow instead of the exact evolution course of the underlying structure, the optimal time window can be determined by either the optimal time window dividing algorithm or by comparing the BICs given by the models with equally divided time windows with different lengths.

*Effects of HRF Delay and Down-sampling on the Proposed Methods*

The effects of the HRF delay on the Granger causality analysis have been discussed by many researches. In this paper, we found that, as the opposite delay increased, the TP rate dropped down and the FP rate rose, which is consistent with the previous results (Smith et al., 2012). In (Deshpande et al., 2010), the authors convolved the HRF with the local field potentials (LFP) recorded from a macaque, and found that even if the HRF delay opposed the underlying neuronal delays is as long as 2.5 seconds, the minimum detectable neuronal delay will still be on the order of a hundred milliseconds. Most recently, Schippers et al. (2011) conducted another simulation-based investigation for the same issue, and found that Granger causal inference can successfully detect over 80% of the cases when the influences flowing toward a region with a faster hemodynamic delay if the neuronal delays are above 1 s. These results suggest that the Granger causality analysis (GCA) performs well when the HRF delay between regions is short; however, when the HRF delay is long, additional procedures must be taken to minimize the effects of the HRF delay on the results given by the GCA. In this paper, we tried to de-convolute a neuronal signal from a BOLD signal using an advanced Kalman filter (Havlicek et al., 2011), but no significant improvement was observed (data not shown). Assuming that a regional HRF delay can be estimated accurately, the performance of the GCA can be improved by realigning the BOLD signals from two regions to control the HRF delay. Note that, to make this realignment work, the sampling rate of the BOLD signal must be finer than the HRF delay between the two regions of interest. Typically, the TR from an fMRI is around 2 to 3 s for whole brain imaging, and by sacrificing the spatial coverage and spatial resolution, the temporal resolution can be as high as 500 ms (Arichi et al., 2012). Fortunately, the speed of an fMRI has been rapidly increasing (Feinberg and Yacoub, 2012), and a sub-second whole-brain fMRI has already been made available (Feinberg et al., 2010). In fact, the most recent advance in MRI technology has enabled a temporal resolution of as fast as 50 ms (Boyacioglu and Barth, 2012). Meanwhile, the accurate and robust estimation of HRF in a BOLD signal has been a fundamental and hot issue for a long time in the area of fMRI data analysis, and many estimation methods have been proposed, including the Friston et al.'s classical paper (1994) and the most recent development by (Wang et al., 2011), among many others. Therefore, an accurate estimation of the regional HRF and a realignment of the BOLD signal to correct the HRF delay are some of our future aims for a GCA of fMRI data.

*Comparison with Filter-based Approaches*

Considering the regional variation of HRF and physiological noise, we simulated the fMRI time series. Based on this dataset, we compared the performances of the proposed optimal Granger causality (GC) approaches with the classic GC and dual Kalman filter cumulative GC, and discussed the effects of the regional variation of the HRF on the Granger causality analysis. The optimal GC approaches outperformed the other two methods for this simulation. We do not intend to imply through this example that our approaches *must* be better than the dual Kalman filter approach. However, the extension of our GC definition to other approaches handling time-varying dynamics is definitely an interesting and important issue that may provide new insight into GC and time-varying dynamics theories, and is one of our future research aims.

*Precuneus Role as a Hub during a Resting State*

For another resting-state fMRI dataset, the proposed approach succeeded in detecting a number of Granger causal interdependencies, from the precuneus to other brain regions, which cannot be inferred by the classic Granger causality, based on the static MVAR model for the whole BOLD time courses. In particular, a circuit centred at the precuneus to the visual network provided proof of the pivotal role played by the precuneus in visual cognition.

*Possibility of Detecting a Status Change in fMRI Data*

In attention-task fMRI data, our approach with spatio-temporally finer-scale details detected the information transfer flows between the brain areas of rIPS and rTPJ. One possible application for this method is detecting the status change in the data. However, the experimental design used in this study for the attention-task was a mixed blocked/ER design. The stimuli were randomized for each subject and block (see the experimental design). The responses were not required for all stimuli, and were therefore also randomized. Both the randomized stimuli and responses might impact the dynamics of the BOLD signal, as well as the block onset and offset. This could cause the detection of unpredictable change-points within the block. Therefore, the current experimental design is not optimal for this purpose, which may be a separate issue of importance. An additional experimental design and the development of a new method may be required in the future.

*Further Directions in Spatio-temporal Granger Causality Algorithm*

The proposed framework, called spatio-temporal Granger causality, consists of several modules, including change-point detection, parameter estimation, and causality estimation. We emphasize that the current algorithm is not the optimal, since a more sophisticated method for each module may improve the overall performance of the analysis. Our future work will aim at finding better algorithms for a more precise estimation of the global Granger causality under the current framework, using up-to-date approaches for each module and a comparison with the existing algorithms (Cribben et al., 2012; Havlicek et al., 2010; Hemmelmann et al., 2009; Hesse et al., 2003).

## Conclusions

The estimation of Granger causality is heavily influenced by the model used. Our results show that a coarse-grained approach/model may average out the meaningful information, since ‘devil is in the details’. The widely held belief that better statistics (Granger causality) result from a longer recording of a dataset is not always true if the whole long-term time series is incorporated into the coarse-grained MVAR model.

Instead, we suggest that the optimal strategy is to divide a long-term recording into a number of time windows using some optimal BIC-based algorithms. A reliable estimation of the Granger causality by a finer-scale MVAR model both in time and space can be achieved.

We proposed a new framework for inferring the Granger causality between groups of times series by taking the finer-scale details into the MVAR model. Our approach shows power to detect an information transfer between brain regions based on fMRI BOLD signals and to enhance the reliability of the estimation. This idea and approach may give rise to a new angle toward the debate of the reliability of Granger causality for fMRI data, particularly the resting-state fMRI time courses.

**Acknowledgements**

We acknowledge two anonymous reviewers for their insightful comments and suggestions. JF is a Royal Society Wolfson Research Merit Award holder, partially supported by National Centre for Mathematics, Interdisciplinary Sciences (NCMIS) of the Chinese Academy of Sciences and the Key Program of National Natural Science Foundation of China (No. 91230201). QL is partially supported by grant from the National Natural Sciences Foundation of China (No.s 11101429, 11271121, 71171195), Research Fund for the Doctoral Program of Higher Education of China (No. 20114307120019), and the National Basic Research Program of China (No. 2011CB707802). LWL is jointly supported by the Marie Curie International Incoming Fellowship from the European Commission (FP7-PEOPLE-2011-IIF-302421), the National Natural Sciences Foundation of China (No. 61273309), the Foundation for the Author of National Excellent Doctoral Dissertation of PR China (No. 200921), Shanghai Rising-Star Program (No. 11QA1400400), and also by the Laboratory of Mathematics for Nonlinear Science, Fudan University.

**Appendix A: Solution of the time-varying linear regression.**

To build up a theoretical analysis of the Granger causality, we assume that the time series are *generated* by the following the first-order (discrete-time) time-varying multivariate autoregressive (TV-MVAR) model:

$$x^{(t+1)} = a_1(t)x^t + b_1(t)y^t + n(t), t = 1, 2, \cdots, T, \quad \text{(A1)}$$

where $n_t$ is white Gaussian noise statistically independent of $x$ and $y$:

$$En(t) = 0, \qquad E[n(t)n(t')] = \sigma_n^2(t)\delta_{t,t'}.$$

Here, $\delta_{t,t'}$ is the Kronecker delta. Without loss of generality, we can suppose that $x^t$ and $y^t$ are centred, *i.e.*, all means are equal to zeros, and the variances of $x^t$ and $y^t$ both equal to 1, by multiplying coefficients $a_1(t)$ and $b_1(t)$ by their variances, respectively. Moreover, we assume that the correlation between $x^t$ and $y^t$ are stationary, i.e., $E\left(x^t y^t\right) = c$ for a constant $c \in [0,1]$. Thus, we can perform a simple linear transformation to make x and y orthogonal:

$$z^t = \frac{(y^t - cx^t)}{\sqrt{1-c^2}},$$

which implies that $z^t$ has its mean equal to 0 and its variance equal to 1, and is uncorrelated with $x^t$. Thus, (A1) becomes:

$$x^{t+1} = \tilde{a}_1(t)x^t + \tilde{b}_1(t)z^t + n^t$$

with $\tilde{a}_1(t) = a_1(t) + b_1(t)c$, $\tilde{b}_1(t) = b_1(t)\sqrt{1-c^2}$. Hence, we can discuss this problem assuming that $x^t$ and $y^t$ are uncorrelated that will not lose generality. Therefore, in the following, we assume that $x^t$ and $y^t$ are uncorrelated.

Considering the time-varying linear regression system (A1), we estimate the Granger causality with different time-window split. More generally, we consider (2) or (3) to replace the intrinsic system. To estimate the **theoretical** values of the time-varying Granger causalities by averaging or cumulating as mentioned in the main text, first, we are to estimate the parameters $\overline{a}_1^{S_1}(k)$ and $\overline{b}_1^{(S_1)}(k)$ by minimizing the following residual square errors across the whole time interval:

$$\begin{aligned}
&\sum_{t=1}^{T} E\left[\left(x^{t+1}-\overline{a}_1^{S_1}(k)x^t-\overline{b}_1^{S_1}(k)y^t\right)^2\right] \\
&=\sum_{t=1}^{T} E\left\{\left[\left(a_1(t)-\overline{a}_1^{S_1}(k)\right)x^t+\left(b_1(t)-\overline{b}_1^{S_1}(k)\right)x^t+n^t\right]^2\right\} \\
&=\sum_{k=1}^{m}\sum_{t=t_{k-1}}^{t_k-1} E\left\{\left[\left(a_1(t)-\overline{a}_1^{S_1}(k)\right)x^t+\left(b_1(t)-\overline{b}_1^{S_1}(k)\right)y^t+n^t\right]^2\right\} \\
&=\sum_{k=1}^{m}\sum_{t=t_{k-1}}^{t_k-1}\left[\left(a_1(t)-\overline{a}_1^{S_1}(k)\right)^2+\left(b_1(t)-\overline{b}_1^{S_1}(k)\right)^2+\sigma_n^2(t)\right]
\end{aligned}$$

which is equivalent to a series of minimisation problems:

$$\min_{\overline{a}_1^{S_1}(k)}\sum_{t=t_{k-1}}^{t_k-1}\left(a_1(t)-\overline{a}_1^{S_1}(k)\right)^2,\quad \min_{\overline{a}_1^{S_1}(k)}\sum_{t=t_{k-1}}^{t_k-1}\left(b_1(t)-\overline{b}_1^{S_1}(k)\right)^2$$

for all $k=1,\cdots,m.$ It can be seen that the (expectation of) the solution should be

$$\overline{a}_1^{S_1}(k)=\frac{1}{t_k-t_{k-1}}\sum_{t=t_{k-1}}^{t_k-1}a_1(t),\ \ \overline{b}_1^{S_1}(k)=\frac{1}{t_k-t_{k-1}}\sum_{t=t_{k-1}}^{t_k-1}b_1(t),\text{ for all } k. \tag{A2}$$

**Appendix B: Monotonicity of the Ganger causalities of TV-MVAR models**

*Monotonicity of cumulative Granger causality.*

By the estimation of the coefficients, Eq. (A2), the cumulative Granger causality with the given time window lengths can be estimated as:

$$F_{Y\to X}^{(c,S_1)}=\log\left(\frac{U_{S_1}+\sum_{t=1}^{T}(b_1(t))^2+\sum_{t=1}^{T}\sigma_n^2(t)}{U_{S_1}+V_{S_1}+\sum_{t=1}^{T}\sigma_n^2(t)}\right),$$

where

$$\mathrm{U}_{S_1}=\sum_{k=1}^{m}\sum_{t=t_{k-1}}^{t_k-1}\left(a_1(t)-\overline{a}_1^{S_1}(k)\right)^2,\ \mathrm{V}_{S_1}=\sum_{k=1}^{m}\sum_{t=t_{k-1}}^{t_k-1}\left(b_1(t)-\overline{b}_1^{S_1}(k)\right)^2.$$

We have the following result.

*Theorem B1.For two change-point sets* $S_1$ *and* $S_2$*, if* $S_1\subseteq S_2$*, then*

$$F_{Y\to X}^{(c,S_1)}\le F_{Y\to X}^{(c,S_2)}$$

*Proof.* Let $S_1$ be composed of the following integer series:

$$1 = t_0 < t_1 < \cdots < t_{m-1} < t_m = T + 1$$

Since the increasing integer series $S_2$ contains $S_1$, we can denote $S_2$ as follows;

$$\begin{aligned} 1 = (t_0 =) t_1^1 < t_1^2 < \cdots < t_1^{n_1} < t_1^{n_1+1} = t_2^1 (= t_1) \\ < \cdots < t_{m-1}^{n_{m-1}+1} = t_m^1 (= t_{m-1}) < t_m^2 < \cdots < t_m^{n_m} (= t_m) = T + 1. \end{aligned}$$

In other words, in each time interval defined by $S_1$, for instance, from $t_{k-1}$ to $t_k$, we denote $t_k^1 < t_k^2 < \cdots < t_k^{n_k} < t_k^{n_k+1}$ as the integers in $S_2$, which are located between $t_{k-1}$ and $t_k$. For simplicity, we let $t_k^{n_k+1} = t_{k+1}^1$. Then, the TV-MVAR model with respect to $S_2$ can be formulated as

$$x^{t+1} = \bar{a}_1^{S_2}(k,q) x^t + \bar{b}_1^{S_2}(k,q) x^t + \tilde{n}(t),$$

$$t_k^q \le t < t_k^{q+1}, q = 1, \cdots, n_k, k = 1, \cdots, m. \qquad (B1)$$

First, we are to prove that $U_{S_1} \ge U_{S_2}$ and $V_{S_1} \ge V_{S_2}$ that are essentially the same. So, we need to prove one of them, for instance, $V_{S_1} \ge V_{S_2}$.

In fact, we rewrite $V_{S_2}$ as follows:

$$V_{S_2} = \sum_{k=1}^{|S_1|} \sum_{q=1}^{n_k} \sum_{t=t_k^q}^{t_k^{q+1}-1} \left( b_1(t) - \bar{b}_1^{S_2}\left(\tau(t_k^q)\right) \right)^2$$

where $\tau(t_k^q)$ denotes the order of $t_k^q$ in the ordered integer set $S_2$.

Thus, it is sufficient to show that in each time window of $S_1$, it holds that

$$\sum_{t=t_{k-1}}^{t_k-1} \left( b_1(t) - \bar{b}_1^{S_1}(k) \right)^2 \ge \sum_{q=1}^{n_k} \sum_{t=t_k^q}^{t_k^{q+1}-1} \left( b_1(t) - \bar{b}_1^{S_2}\left(\tau(t_k^q)\right) \right)^2.$$

We note that

$$\sum_{t=t_{k-1}}^{t_k-1} \left( b_1(t) - \bar{b}_1^{S_1}(k) \right)^2 = \sum_{t=t_{k-1}}^{t_k-1} [b_1(t)]^2 - (t_k - t_{k-1})[\bar{b}_1^{S_1}(k)]^2,$$

$$\sum_{q=1}^{n_k} \sum_{t=t_k^q}^{t_k^{q+1}-1} \left( b_1(t) - \bar{b}_1^{S_2}\left(\tau(t_k^q)\right) \right)^2 = \sum_{t=t_{k-1}}^{t_k-1} b_1^2(t) - \sum_{q=1}^{n_k} \left( t_k^{q+1} - t_k^q \right) \left( \bar{b}_1^{S_2}\left(\tau(t_k^q)\right) \right)^2$$

and

$$\bar{b}_1^{S_1}(k) = \frac{1}{t_k - t_{k-1}} \sum_{q=1}^{n_k} \left(t_k^{q+1} - t_k^q\right) \bar{b}_1^{S_2}\left(\tau(t_k^q)\right), \sum_{q=1}^{n_k} \left(t_k^{q+1} - t_k^q\right) = t_k - t_{k-1}.$$

Hence,

$$\begin{aligned}
&(t_k - t_{k-1})[\bar{b}_1^{S_1}(k)]^2 \\
&= (t_k - t_{k-1}) \left[ \frac{1}{t_k - t_{k-1}} \sum_{q=1}^{n_k} \left(t_k^{q+1} - t_k^q\right) \bar{b}_1^{S_2}\left(\tau\left(t_k^q\right)\right) \right]^2 \\
&\le (t_k - t_{k-1}) \frac{1}{t_k - t_{k-1}} \sum_{q=1}^{n_k} \left(t_k^{q+1} - t_k^q\right) [\bar{b}_1^{S_2}\left(\tau\left(t_k^q\right)\right)]^2 \\
&= \sum_{q=1}^{n_k} \left(t_k^{q+1} - t_k^q\right) [\bar{b}_1^{S_2}\left(\tau\left(t_k^q\right)\right)]^2
\end{aligned} \tag{B2}$$

owing to the well-known fact that the weighted algebraic average is less than the square average with the weighting. This implies $\mathrm{V}_{S_1} \ge \mathrm{V}_{S_2}$. So, it is with $\mathrm{U}_{S_1} \ge \mathrm{U}_{S_2}$.

From $\mathrm{V}_{S_1} \ge \mathrm{V}_{S_2}$, we immediately have

$$\frac{\sum_{t=1}^{T} (b_1(t))^2 + \sum_{t=1}^{T} \sigma_n^2(t)}{\mathrm{V}_{S_1} + \sum_{t=1}^{T} \sigma_n^2(t)} \le \frac{\sum_{t=1}^{T} (b_1(t))^2 + \sum_{t=1}^{T} \sigma_n^2(t)}{\mathrm{V}_{S_2} + \sum_{t=1}^{T} \sigma_n^2(t)}$$

Combined by $\mathrm{U}_{S_1} \ge \mathrm{U}_{S_2}$, we can derive

$$\frac{\mathrm{U}_{S_1} + \sum_{t=1}^{T} (b_1(t))^2 + \sum_{t=1}^{T} \sigma_n^2(t)}{\mathrm{U}_{S_1} + \mathrm{V}_{S_1} + \sum_{t=1}^{T} \sigma_n^2(t)} \le \frac{\mathrm{U}_{S_2} + \sum_{t=1}^{T} (b_1(t))^2 + \sum_{t=1}^{T} \sigma_n^2(t)}{\mathrm{U}_{S_2} + \mathrm{V}_{S_2} + \sum_{t=1}^{T} \sigma_n^2(t)}$$

This means $F_{Y \to X}^{(c,S_1)} \le F_{Y \to X}^{(c,S_2)}$. From (B2), one can see that the inequality holds if and only if

$$\bar{b}_1^{S_2}\left(\tau(t_k^q)\right) = \bar{b}_1^{S_1}(k) \tag{B3}$$

holds for all $q = 1, \cdots, n_k$ and all $k$. □

From Theorem B1 and its proof, in particular Eq. (B3) as the sufficient and necessary condition for $F_{Y \to X}^{(c,S_1)} = F_{Y \to X}^{(c,S_2)}$. We immediately have the upper and lower bounds of the cumulative Granger causality.

*Corollary B2. Let* $S_0 = \{1, T+1\}$ *and* $S_*$ *be the ordered time point set that exactly*

*comprise of the change-points in the TV- MVAR. Then, for any ordered time point set S, we have*

$$F_{Y\to X}^{(c,S_0)} \le F_{Y\to X}^{(c,S)} \le F_{Y\to X}^{(c,S_*)}$$

This corollary shows that the static (classic) Granger causality actually is the lower-bound of the cumulative Granger causality. And, if the time series are exactly generated by TV-MVAR (A1) with the change-point set $S_*$, the cumulative Granger causality based on it is the upper bounds of all.

We should point out that Theorem B1 holds under the condition that one switching time set is contained in the other. It does not imply that the more change-points are, the larger cumulative Granger causality it will have.

*Conjecture 1.If* $|S_2| \ge |S_1|$, *then* $F_{Y\to X}^{(c,S_1)} \le F_{Y\to X}^{(c,S_2)}$.

We claim that this conjecture is ***not true*** by a simple counter example. Let $T$ = 6, $S_1 = \{1,4,7\}$, $S_2 = \{1,3,5,7\}$, and $S = \{1,7\}$. We suppose that the data is produced by the model (5) with a constant $a_1$ and $b_1(t)$ is periodic with a period 2, *i.e*, $b_1(1) = b_1(3) = b_1(5)$ and $b_1(2) = b_1(4) = b_1(6)$. But the two values do not equal pair-wisely. It is clear that for $S_2$, the parameters can be estimated as

$$\bar{b}_1^{S_2}(1) = \bar{b}_1^{S_2}(2) = \bar{b}_1^{S_2}(3) = \frac{1}{2}\left(b_1(1) + b_1(2)\right),$$

which equals to the whole average

$$\bar{b}_1 = \frac{1}{6}\left(b_1(1) + b_1(2) + b_1(3) + b_1(4) + b_1(5) + b_1(6)\right).$$

So, the corresponding Granger Causality with $S_2$ can be estimated equal to that of the static MVAR model (the change-point is composed of $S$), i.e., $F_{Y\to X}^{(c,S_2)} = F_{Y\to X}^{(c,S)}$.

Noting that $F_{Y\to X}^{(c,S)} < F_{Y\to X}^{(c,S_1)}$, where the strict inequality is because of $\bar{b}_1^{S_1}(1) = \frac{1}{3}\left(b_1(1) + b_1(2) + b_3(2)\right) \ne \bar{b}_1$. So, we have $F_{Y\to X}^{(c,S_2)} < F_{Y\to X}^{(c,S_1)}$ despite $|S_2| > |S_1|$.

Let us consider a numerical example with $a_1(t) = 1$, $\sigma_n^1(t) = 1$ for all $t$, and

$b_1(1)=b_1(3)=b_1(5)=1$ and $b_1(2)=b_1(4)=b_1(6)=0$. Direct calculations lead that

$$F_{Y\to X}^{(\mathrm{c},\mathrm{S}_1)}=\log\left(\frac{\sum_{t=1}^{6}(b_1(t))^2+6}{\sum_{k=1}^{2}\sum_{t=3(k-1)+1}^{3k}\left(b_1(t)-\overline{b}_1^{i}(k)\right)^2+6}\right)$$
$$=\log\frac{6+3}{8/9+6}=\log\frac{81}{62}.$$

However,

$$F_{Y\to X}^{(\mathrm{c},\mathrm{S}_2)}=\log\left(\frac{\sum_{t=1}^{6}(b_1(t))^2+6}{\sum_{k=1}^{3}\sum_{t=2(k-1)+1}^{2k}\left(b_1(t)-\overline{b}_1^{i}(k)\right)^2+6}\right)$$
$$=\log\frac{6+3}{3/2+6}=\log\frac{6}{5}<\log\frac{81}{62}=F_{Y\to X}^{(\mathrm{c},\mathrm{S}_1)}.$$

*Monotonicity of average Granger causality.*

Another approach to estimate the Granger causality of TV-MVAR model is to estimate the Granger causality at each time windows (between switching) can average them according to the length of each time window. Recall $S_1=\{1=t_0<t_1<\cdots<t_{m-1}<t_m=T+1\}$ as an increasing integer sequence that denotes the change-point and the TV-MVAR model as

$$x^{t+1}=\overline{a}_1^{S_1}(k)x^t+\overline{b}_1^{S_1}(k)x^t+\breve{n}(t),t_{k-1}\le t<t_k,k=1,\cdots,m \quad \text{(B4)}$$

At each time window, the Granger causality can be estimated as

$$F_{Y\to X}^{(k,S_1)}=\log\left(\frac{U_k+\sum_{t=t_{k-1}}^{t_k-1}(b_1(t))^2+\sum_{t=t_{k-1}}^{t_k-1}\sigma_n^2(t)}{U_k+V_k+\sum_{t=t_{k-1}}^{t_k-1}\sigma_n^2(t)}\right)$$

With

$$\mathrm{U}_k=\sum_{t=t_{k-1}}^{t_k-1}\left(a_1(t)-\overline{a}_1^{S_1}(k)\right)^2,\mathrm{V}_k=\sum_{t=t_{k-1}}^{t_k-1}\left(b_1(t)-\overline{b}_1^{S_1}(k)\right)^2.$$

Then, we estimate the Granger Causality by the TV-MVAR model (B4) as follows:

$$F_{Y\to X}^{(a,S_1)}=\frac{1}{T}\sum_{k=1}^{m}F_{Y\to X}^{(k,S_1)}(t_k-t_{k-1}),$$

named the average Granger causality, the weighted average according to the lengths of the time windows. To investigate the relationship between the average Granger causality and the fineness of temporal resolution, we need the following lemma:

*Lemma B3. For any positive integer T, any m real constants* $[c_t]_{t=1}^T$ *, any T nonnegative constants* $[p_t]_{t=1}^T$ *with* $\sum_{t=1}^T p_t = 1$ *and any positive constants* $\left[\sigma_t^2\right]_{t=1}^T$*, we have the following inequality*

$$\log\left(\frac{\sum_{t=1}^T c_t^2 p_t + \sigma^2}{\sum_{t=1}^t (c_t - \bar{c})^2 p_t + \sigma^2}\right) \le \sum_{t=1}^t p_t \log\left(\frac{c_t^2 + \sigma_t^2}{\sigma_t^2}\right)$$

*where* $\bar{c} = \sum_{t=1}^T p_t c_t$ *and* $\sigma^2 = \sum_{t=1}^T p_t \sigma_t^2$.

*Proof.* Let us consider the following function with respect to $C = [c_t]_{t=1}^T$ and $\Sigma = \left[\sigma_t^2\right]_{t=1}^T$

$$V = \log \frac{\sigma^2 + c^2}{\sigma^2 + v} - \sum_{t=1}^T p_t \log\left(\frac{\sigma_t^2 + c_t^2}{\sigma_t^2}\right)$$

with

$$c^2 = \sum_{t=1}^T c_t^2 p_t, v = \sum_{t=1}^T [c(t) - \bar{c}]^2 p_t = c^2 - (\bar{c})^2.$$

We make minor modifications on the problem according to the following three facts.

First, noting

$$v = \sum_{t=1}^T [c_t - \bar{c}]^2 p_t = \sum_{t=1}^T [c_t)]^2 p_t - \left(\sum_{t=1}^T c_t p_t\right)^2 \ge \sum_{t=1}^T |c_t|^2 p_t - \left(\sum_{t=1}^T |c_t| p_t\right)^2,$$

if we replace $c_t$ by $|c_t|$ in *V*, which is denoted by $\hat{V}$, then we have $V \le \hat{V}$.

Therefore, without loss of generality, it is sufficient to prove $\hat{V} \le 0$ by considering the case that all $c_t$ are nonnegative.

Second, considering

$$\left.\frac{\partial V}{\partial c_t}\right|_{c_t=0} = \frac{2\bar{c}p_t}{\sigma^2 + v},$$

which is positive in the case that there is at least one positive $c_t$ with nonzero $p_t$. So, it is sufficient to consider the case that all $c_t$ are positive;

Third, it holds

$$\sum_{t=1}^{T} p_t \log\left(\sigma_t^2\right) \le \log(\sigma^2)$$

owing to the Jensen's inequality.

In summary, we can consider the following function instead of *V*:

$$\tilde{V}\left(y, v, \Sigma\right) = \log \frac{\sigma^2 + c^2}{\sigma^2 + v} - \sum_{t=1}^{T} p_t \log\left(\frac{\sigma_t^2 + y_t}{\sigma^2}\right).$$

Letting $y_t = (c_t)^2$, owing to the fact that all $c_t$ are nonnegative, $y = [y_t]$, with fixed $c^2$ and $\sigma^2$, we are going to show that $\tilde{V}$ is nonnegative by considering the following maximization problem:

$$\max \tilde{V}(y, v, \Sigma)$$

$$\text{s.t. } \sum_{t=1}^{T} y_t p_t = c^2, \sum_{t=1}^{T} \sqrt{y_t}\, p_t = \sqrt{c^2 - v}, \sum_{t=1}^{T} p_t \sigma_t^2 = \sigma^2, \sigma_t^2 > 0, y_t > 0, \forall t \qquad \text{(B5)}$$

To solve (B5), we introduce the following auxiliary Lagrange function:

$$L\left(y, v, \Sigma, \lambda, \mu, \gamma\right)$$
$$= \tilde{V}\left(y, v, \Sigma\right) + \lambda\left(\sum_{t=1}^{T} y_t p_t - b^2\right) + \mu\left(\sum_{t=1}^{T} \sqrt{y_t}\, p_t - \sqrt{c^2 - v}\right) + \gamma\left(\sum_{t=1}^{T} \sigma_t^2 p_t - \sigma^2\right)$$

By the Karush-Kuhn-Tucker conditions, the necessary conditions of the minimum of (B5) include:

$$\frac{\partial L}{\partial y_t} = p_t\left[-\frac{1}{\left(\sigma_t^2 + y_t\right)} + \lambda + \frac{\mu}{2\sqrt{y_t}}\right] = 0,$$

$$\frac{\partial L}{\partial v} = -\frac{1}{\sigma^2 + v} + \frac{\mu}{2\sqrt{c^2 - v}} = 0,$$

$$\frac{\partial L}{\partial (\sigma_t^2)} = p_t\left[-\frac{1}{\left(\sigma_t^2 + y_t\right)} + \gamma\right] = 0$$

This leads (i) $p_t = 0$ or (ii) $\sigma_t^2 + y_t = 1/\gamma$ and

$$y_t = \left[\frac{\mu}{2(\gamma-\lambda)}\right]^2. \tag{B6}$$

In other words, for these $y_t$ with nonzero $p_t$, if we are to solve $y_t$ from the above equalities as a function with respect to $\sigma, \lambda, \mu, \gamma, c^2$ and *v*, which are independent of the index *t*, then we can only have one expression from (B6). It should be emphasized that we are not solving the values of $y_t$ but its expression with respect to the *t*-independent quantities, $\sigma, \lambda, \mu, \gamma, c^2$ and *v*. Therefore, the possible minimum points of $R(y,v)$ only has one single value of $y_t$. So it is with

$$\sigma_t^2 = \frac{1}{\gamma} - \left[\frac{\mu}{2(\gamma-\lambda)}\right]^2$$

It can be seen that if $y_t$ and $\sigma_t^2$ can only have a single value respectively, then $\tilde{V}(y,v,\Sigma)=0$. So, the maximum of $\tilde{V}(y,v,\Sigma)$ is zero. Hence, the intrinsic $V$ has its maximum equal to zero. Therefore, lemma B2 is proved and the equality holds if and only if $y_t$ and $\sigma_t^2$ can only have a single value respectively. □

*Theorem B4. Let* $S_1$ *and* $S_2$ *be two sequences of increasing integers. If* $S_1 \subseteq S_2$, *then*

$$F_{Y\to X}^{(a,S_1)} \le F_{Y\to X}^{(a,S_2)}.$$

*Proof.* We denote the sets $S_1$ *and* $S_2$ by the symbols as in the proof of Theorem B1. According to (B1), the Granger causality at the time window of $S_2$, the *k*-th window of $S_1$ and the *q*-th can be written as:

$$F_{Y\to X}^{(k,q,S_2)} = \log\left(\frac{U_{k,q} + \sum_{t=t_k^q}^{t_k^{q+1}-1}(b_1(t))^2 + \sum_{t=t_k^q}^{t_k^{q+1}-1}\sigma_n^2(t)}{U_{k,q} + V_{k,q} + \sum_{t=t_k^q}^{t_k^{q+1}-1}\sigma_n^2(t)}\right)$$

where

$$U_{k,q} = \sum_{t=t_k^q}^{t_k^{q+1}-1} \left(a_1(t) - \overline{a}_1^{S_2}\left(\tau(t_k^q)\right)\right)^2, V_{k,q} = \sum_{t=t_k^q}^{t_k^{q+1}-1} \left(b_1(t) - \overline{b}_1^{S_2}\left(\tau(t_k^q)\right)\right)^2.$$

Then, the average Granger causality with respect to $S_2$ is

$$F_{Y\to X}^{(a,S_2)} = \frac{1}{T}\sum_{k=1}^{|S_1|}\sum_{q=1}^{n_k} F_{Y\to X}^{(k,q,S_2)}(t_k^{q+1} - t_k^q).$$

Note

$$\sum_{q=1}^{n_k} \overline{a}_1^{S_2}\left(\tau\left(t_k^q\right)\right)(t_k^{q+1} - t_k^q) = \overline{a}_1^{S_1}(k), \sum_{q=1}^{n_k} \overline{b}_1^{S_2}\left(\tau\left(t_k^q\right)\right)(t_k^{q+1} - t_k^q) = \overline{b}_1^{S_1}(k),$$

and

$$\sum_{t=t_k^q}^{t_k^{q+1}-1} [a_1(t)]^2 = \sum_{t=t_k^q}^{t_k^{q+1}-1} [a_1(t) - \overline{a}_1^{S_2}\left(\tau\left(t_k^q\right)\right)]^2 + \left[\overline{a}_1^{S_2}\left(\tau\left(t_k^q\right)\right)\right]^2 \left(t_k^{q+1} - t_k^q\right),$$

$$\sum_{t=t_k^q}^{t_k^{q+1}-1} [b_1(t)]^2 = \sum_{t=t_k^q}^{t_k^{q+1}-1} [a_1(t) - \overline{b}_1^{S_2}\left(\tau\left(t_k^q\right)\right)]^2 + \left[\overline{b}_1^{S_2}\left(\tau\left(t_k^q\right)\right)\right]^2 \left(t_k^{q+1} - t_k^q\right).$$

Compared with $F_{Y\to X}^{(a,S_2)}$, we can rewrite the term of $F_{Y\to X}^{(a,S_1)}$, i.e., $F_{Y\to X}^{(k,S_1)}$, as follows:

$$\begin{aligned}
F_{Y\to X}^{(k,S_1)} &= \log\left(\frac{U_k + \sum_{t=t_{k-1}}^{t_k-1}(b_1(t))^2 + \sum_{t=t_{k-1}}^{t_k-1}\sigma_n^2(t)}{U_k + V_k + \sum_{t=t_{k-1}}^{t_k-1}\sigma_n^2(t)}\right) \\
&= \log\Bigg(\sum_{q=1}^{n_k}\left[\overline{a}_1^{S_1}(k) - \overline{a}_1^{S_2}\left(\tau\left(t_k^q\right)\right)\right]^2\left(t_k^{q+1} - t_k^q\right) \\
&\quad + \sum_{q=1}^{n_k}\left[\overline{b}_1^{S_2}\left(\tau\left(t_k^q\right)\right)\right]^2\left(t_k^{q+1} - t_k^q\right) + \sum_{q=1}^{n_k}\varepsilon_n^2(q)\left(t_k^{q+1} - t_k^q\right)\Bigg) \\
&\quad - \log\Bigg(\sum_{q=1}^{n_k}\left[\overline{a}_1^{S_1}(k) - \overline{a}_1^{S_2}\left(\tau\left(t_k^q\right)\right)\right]^2\left(t_k^{q+1} - t_k^q\right) \\
&\quad + \sum_{q=1}^{n_k}\left[\overline{b}_1^{S_1}(k) - \overline{b}_1^{S_2}\left(\tau\left(t_k^q\right)\right)\right]^2\left(t_k^{q+1} - t_k^q\right) + \sum_{q=1}^{n_k}\varepsilon_n^2(q)\left(t_k^{q+1} - t_k^q\right)\Bigg)
\end{aligned}$$

where

$$\varepsilon_n^2(q) = \frac{1}{t_k^{q+1} - t_k^q}\sum_{t=t_k^q}^{t_k^{q+1}-1}\left\{\left[a_1(t) - \overline{a}_1^{S_2}\left(\tau\left(t_k^q\right)\right)\right]^2 + \left[b_1(t) - \overline{b}_1^{S_2}\left(\tau\left(t_k^q\right)\right)\right]^2 + \sigma_n^2(t)\right\}.$$

Since

$$\frac{\sum_{q=1}^{n_k}\left[\overline{a}_1^{S_1}(k)-\overline{a}_1^{S_2}\left(\tau\left(t_k^q\right)\right)\right]^2\left(t_k^{q+1}-t_k^q\right)+\sum_{q=1}^{n_k}\left[\overline{b}_1^{S_2}\left(\tau\left(t_k^q\right)\right)\right]^2\left(t_k^{q+1}-t_k^q\right)+\sum_{q=1}^{n_k}\varepsilon_n^2(q)\left(t_k^{q+1}-t_k^q\right)}{\sum_{q=1}^{n_k}\left[\overline{a}_1^{S_1}(k)-\overline{a}_1^{S_2}\left(\tau\left(t_k^q\right)\right)\right]^2\left(t_k^{q+1}-t_k^q\right)+\sum_{q=1}^{n_k}\left[\overline{b}_1^{S_1}(k)-\overline{b}_1^{S_2}\left(\tau\left(t_k^q\right)\right)\right]^2\left(t_k^{q+1}-t_k^q\right)+\sum_{q=1}^{n_k}\varepsilon_n^2(q)\left(t_k^{q+1}-t_k^q\right)}$$

$$\leq\frac{\sum_{q=1}^{n_k}\left[\overline{b}_1^{S_2}\left(\tau\left(t_k^q\right)\right)\right]^2\left(t_k^{q+1}-t_k^q\right)+\sum_{q=1}^{n_k}\varepsilon_n^2(q)\left(t_k^{q+1}-t_k^q\right)}{\sum_{q=1}^{n_k}\left[\overline{b}_1^{S_1}(k)-\overline{b}_1^{S_2}\left(\tau\left(t_k^q\right)\right)\right]^2\left(t_k^{q+1}-t_k^q\right)+\sum_{q=1}^{n_k}\varepsilon_n^2(q)\left(t_k^{q+1}-t_k^q\right)}$$

Theorem B3 can be derived by directly employing Lemma B3. In addition, the inequality holds if and only if $\overline{b}_1^{S_2}\left(\tau\left(t_k^q\right)\right)$ and $\varepsilon_n^2(q)$ can only pick values independent of the index $q$ (but possibly depending on the index $k$), respectively. □

Similar to Corollary B2, from Theorem B4 and its proof, in particular the sufficient and necessary condition for $F_{Y\to X}^{(a,S_1)}=F_{Y\to X}^{(a,S_2)}$. We immediately have the upper and lower bounds of the cumulative Granger causality.

*Corollary B5. Let* $S_0=\{1,T+1\}$ *and* $S_*$ *be the ordered time point set that exactly comprise of the change-points in the TV-MVAR model. Then, for any ordered time point set S, we have*

$$F_{Y\to X}^{(a,S_0)}\leq F_{Y\to X}^{(a,S)}\leq F_{Y\to X}^{(a,S_*)}.$$

This corollary shows that the static (classic) Granger causality actually is the lower-bound of the average Granger causality. And, if the time series are exactly generated by TV-MVAR (A1) with the change-point set $S_*$, the average Granger causality based on it is the upper bounds of all.

We should also emphasize that the following conjecture is ***not true***.

*Conjecture B2. If* $|S_2|\geq|S_1|$, *then* $F_{Y\to X}^{(a,S_1)}\leq F_{Y\to X}^{(a,S_2)}$.

That is to say, the average Granger causality is monotonic with respect to the containing relation between the change-point set, but not monotonic with respect to the size of the change-point sets. A counter-example can be easily established by the same way as in Remark 1.

**Appendix C: Comparison between cumulative and average Ganger causalities**

*Magnitude comparison.*

Actually, the two sorts of Granger causalities of the TV-MVAR model do not have definite magnitude relation. First, we show in the following theorem, the relation that cumulative Granger causality is greater than the average Granger causality with the same change-point set is **conditional**.

***Theorem C1.*** *Let S be a sequence of increasing integers. If the following quantity*

$$\frac{1}{t_k - t_{k-1}}\left[\sum_{t=t_{k-1}}^{t_k-1}\left(a_1(t)-\overline{a}_1(k)\right)^2+\sum_{t=t_{k-1}}^{t_k-1}\left(b_1(t)-\overline{b}_1(k)\right)^2+\sum_{t=t_{k-1}}^{t_k-1}\sigma_n^2(t)\right]$$

*with*

$$\overline{a}_1(k)=\frac{1}{t_k-t_{k-1}}\sum_{t=t_{k-1}}^{t_k-1}a_1(t),\overline{b}_1(k)=\frac{1}{t_k-t_{k-1}}\sum_{t=t_{k-1}}^{t_k-1}b_1(t)$$

*is independent of the index k, then*

$$F_{Y\to X}^{(a,S)}\leq F_{Y\to X}^{(c,S)}.$$

*Proof.* Let $S=\{1=t_0<t_1<\cdots<t_{m-1}<t_m=T+1\}$. And, with the same notations we used above, we have

$$F_{Y\to X}^{(c,S)}=\log\left(\frac{\sum_{k=1}^{m}\sum_{t=t_{k-1}}^{t_k-1}\left(a_1(t)-\overline{a}_1(k)\right)^2+\sum_{t=1}^{T}(b_1(t))^2+\sum_{t=1}^{T}\sigma_n^2(t)}{\sum_{k=1}^{m}\sum_{t=t_{k-1}}^{t_k-1}\left(a_1(t)-\overline{a}_1(k)\right)^2+\sum_{k=1}^{m}\sum_{t=t_{k-1}}^{t_k-1}\left(b_1(t)-\overline{b}_1(k)\right)^2+\sum_{t=1}^{T}\sigma_n^2(t)}\right),$$

and

$$F_{Y\to X}^{(a,S)}=\frac{1}{T}\sum_{k=1}^{m}(t_k-t_{k-1})\log\left(\frac{\sum_{t=t_{k-1}}^{t_k-1}\left(a_1(t)-\overline{a}_1(k)\right)^2+\sum_{t=t_{k-1}}^{t_k-1}(b_1(t))^2+\sum_{t=t_{k-1}}^{t_k-1}\sigma_n^2(t)}{\sum_{t=t_{k-1}}^{t_k-1}\left(a_1(t)-\overline{a}_1(k)\right)^2+\sum_{t=t_{k-1}}^{t_k-1}\left(b_1(t)-\overline{b}_1(k)\right)^2+\sum_{t=t_{k-1}}^{t_k-1}\sigma_n^2(t)}\right)$$

Let $\overline{b}_1=\frac{1}{T}\sum_{t=1}^{T}b_1(t)$. Thus, we can rewrite them as

$$F_{Y\to X}^{(c,S)}$$

$$=\log\left(\frac{\sum_{k=1}^{m}\sum_{t=t_{k-1}}^{t_k-1}(a_1(t)-\overline{a}_1(k))^2+\sum_{t=1}^{m}(\overline{b}_1(k))^2(t_k-t_{k-1})+\sum_{t=1}^{m}\sum_{t=t_{k-1}}^{t_k-1}(b_1(t)-\overline{b}_1(k))^2+\sum_{t=1}^{T}\sigma_n^2(t)}{\sum_{k=1}^{m}\sum_{t=t_{k-1}}^{t_k-1}(a_1(t)-\overline{a}_1(k))^2+\sum_{k=1}^{m}\sum_{t=t_{k-1}}^{t_k-1}(b_1(t)-\overline{b}_1(k))^2+\sum_{t=1}^{T}\sigma_n^2(t)}\right)$$

and

$$F_{Y\to X}^{(a,S)} = \frac{1}{T}\sum_{k=1}^{m}(t_k - t_{k-1})\times$$

$$\log\left(\frac{\sum_{t=t_{k-1}}^{t_k-1}(a_1(t)-\bar{a}_1(k))^2 + \sum_{t=t_{k-1}}^{t_k-1}(b_1(t)-\bar{b}_1(k))^2 + (\bar{b}_1(k))^2(t_k - t_{k-1}) + \sum_{t=t_{k-1}}^{t_k-1}\sigma_n^2(t)}{\sum_{t=t_{k-1}}^{t_k-1}(a_1(t)-\bar{a}_1(k))^2 + \sum_{t=t_{k-1}}^{t_k-1}(b_1(t)-\bar{b}_1(k))^2 + \sum_{t=t_{k-1}}^{t_k-1}\sigma_n^2(t)}\right).$$

In addition, letting

$$\theta_k = \frac{1}{t_k - t_{k-1}}\left[\sum_{t=t_{k-1}}^{t_k-1}(a_1(t)-\bar{a}_1(k))^2 + \sum_{t=t_{k-1}}^{t_k-1}(b_1(t)-\bar{b}_1(k))^2 + \sum_{t=t_{k-1}}^{t_k-1}\sigma_n^2(t)\right],$$

$$\alpha_k = \frac{1}{t_k - t_{k-1}}\sum_{t=t_{k-1}}^{t_k-1}(\bar{b}_1(k))^2, p_k = \frac{(t_k - t_{k-1})}{T}$$

due to the condition, $\theta_k$ is independent of $k$, which is denoted by $\theta$. Thus, we have

$$F_{Y\to X}^{(c,S)} = \log\left(\frac{\sum_{t=1}^{m}\alpha_k p_k + \sum_{k=1}^{m}\theta_k p_k}{\sum_{k=1}^{m}\theta_k p_k}\right) = \log\left(\frac{\sum_{t=1}^{m}\alpha_k p_k + \theta}{\theta}\right),$$

and

$$F_{Y\to X}^{(a,S)} = \sum_{k=1}^{m} p_k \log\left(\frac{\alpha_k + \theta_k}{\theta_k}\right) = \sum_{k=1}^{m} p_k \log\left(\frac{\alpha_k + \theta}{\theta}\right).$$

Thus, we can conclude $F_{Y\to X}^{(a,S)} \le F_{Y\to X}^{(c,S)}$, owing to the Jensen's inequality. This completes the poof. □

From Theorem C1, if the TV-MVAR system is time-varying with the segments well known, which implies that at each time window, the system is static, we can conclude that the average Grange causality is smaller than the cumulative Granger causality.

On the other hand, if the condition in Theorem C1 is not satisfied, then it will not be surprising that $F_{Y\to X}^{(a,S)} > F_{Y\to X}^{(c,S)}$ holds. Here is a counter-example. A special situation is to solve the time-varying regression model (A1) as a static one, i.e., taking all time points as the change-point set, i.e., $S = \{1,\cdots,T+1\}$. By a proper transformation, we can still let the variances of $y$ and $x$ equal to 1 for all time. Let $a_1(t) = 0$ for all $t$. But the variance of the noise may be time-varying. The defined Granger causalities become:

$$F_{Y\to X}^{S}=\log\left(\frac{\frac{1}{T}\sum_{t=1}^{T}\sigma_n^2(t)+\frac{1}{T}\sum_{t=1}^{T}[b_1(t)]^2}{\frac{1}{T}\sum_{t=1}^{T}\sigma_n^2(t)}\right),\hat{F}_{Y\to X}^{S}=\frac{1}{T}\sum_{t=1}^{T}\log\left(\frac{\sigma_n^2(t)+[b_1(t)]^2}{\sigma_n^2(t)}\right).$$

Pick $T=3, b_1=1, b_2=\sqrt{2}, b_3=\sqrt{3}, \sigma_1=\sqrt{3}, \sigma_2=\sqrt{2}, \sigma_3=1$. Then,

$$F_{Y\to X}=\log 2<\hat{F}_{Y\to X}=\frac{1}{3}\left[\log\left(\frac{4}{3}\right)+\log(2)+\log(4)\right].$$

*Comparison between the asymptotic square moments under null hypothesis.*

For simplicity, we suppose that the time-varying system (A1) is a switching system with equal-length time windows and the segment points are exactly known. The general case will be treated in our future paper. Thus, (A1) becomes a series of static linear system as follows:

$$x^{t+1}=\bar{a}_1(k)x^t+\bar{b}_1(k)y^t+n(t), t_k\le t<t_{k+1}, k=1,2,\cdots m, \tag{C1}$$

Here, $t_{k+1}-t_k=\frac{n}{m}$ for all $k$. Under the null hypothesis, namely, the coefficients $b_1(t)=0$ hold for all *t*, (C1) becomes

$$x^{t+1}=\bar{a}_1(k)x^t+\tilde{n}(t), t_k\le t<t_{k+1}, k=1,2,\cdots m. \tag{C2}$$

Their residual squared errors at each time window are

$$RSS_1(k)=\sum_{t=t_k}^{t_{k+1}-1}n^2(t) \quad \text{and} \quad RSS_0(k)=\sum_{t=t_k}^{t_{k+1}-1}\tilde{n}^2(t).$$

Then, the CGS and AGC can be formulated as follows respectively:

$$F_{Y\to X}^{c}=\log\left(\frac{\sum_{k=1}^{m}RSS_0(k)}{\sum_{k=1}^{m}RSS_1(k)}\right)=\log\left(1+\frac{\sum_{k=1}^{m}[RSS_0(k)-RSS_1(k)]}{\sum_{k=1}^{m}RSS_1(k)}\right)\sim\frac{\sum_{k=1}^{m}[RSS_0(k)-RSS_1(k)]}{\sum_{k=1}^{m}RSS_1(k)}$$

$$F_{Y\to X}^{a}=\frac{1}{m}\sum_{k=1}^{m}\log\left(\frac{RSS_0(k)}{RSS_1(k)}\right)=\frac{1}{m}\sum_{k=1}^{m}\log\left(1+\frac{RSS_0(k)-RSS_1(k)}{RSS_1(k)}\right)\sim\frac{1}{m}\sum_{k=1}^{m}\frac{RSS_0(k)-RSS_1(k)}{RSS_1(k)}$$

as *n* goes to infinity. Therefore, the cumulative Granger causality converges the following in distribution:

$$F_{Y\to X}^{c}\times\frac{n-2m-1}{m}\approx\frac{\sum_{k=1}^{m}[RSS_0(k)-RSS_1(k)]}{\sum_{k=1}^{m}RSS_1(k)}\times\frac{n-2m-1}{m}\sim F(m,n-2m-1)$$

and the average Granger causality converges to:

$$F_{Y\to X}^{a}\times\frac{n/m-3}{1}\approx\frac{1}{m}\sum_{k=1}^{m}\frac{RSS_0(k)-RSS_1(k)}{RSS_1(k)}\times\frac{n/m-3}{1}$$

with each

$$\frac{RSS_0(k)-RSS_1(k)}{RSS_1(k)}\times\frac{n/m-3}{1}\sim F(1,\frac{n}{m}-3)$$

So, their asymptotic expectations are

$$E\left(F_{Y\to X}^{c}\right)\approx\frac{m}{n-2m-1}\frac{n-2m-1}{n-2m-3}=\frac{1}{\frac{n}{m}-2-\frac{3}{m}},E\left(F_{Y\to X}^{a}\right)\approx\frac{\frac{n}{m}-3}{\frac{n}{m}-5}\frac{1}{\frac{n}{m}-3}=\frac{1}{\frac{n}{m}-5},$$

as $n\to\infty$.

The dominant converge rates are same, equivalently $\frac{m}{n}$. Then, let us take a look at their asymptotic square moments. By the square moment of the $F$-distribution and simple algebras, we have

$$E\left\{\left[F_{Y\to X}^{c}\right]^2\right\}\approx\left(1+\frac{2}{m}\right)\left(\frac{m}{n}\right)^2.$$

As for the AGC, with $f_k=\frac{RSS_0(k)-RSS_1(k)}{RSS_1(k)}\sim F(1,\frac{n}{m}-3)$, we have

$$E\left\{\left[F_{Y\to X}^{a}\right]^2\right\}\cong\left(\frac{1}{n/m-3}\right)^2E\left\{\frac{1}{m}\sum_{k=1}^{m}f_k\right\}^2\leq\left(\frac{1}{n/m-3}\right)^2\left(\frac{1}{m}\right)\sum_{k=1}^{m}E\{f_k\}^2$$

With

$$E\left\{f_k^2\right\}\cong 3\left(\frac{\frac{n}{m}-3}{\frac{n}{m}-5}\right)^2,$$

which implies

$$E\left\{\left[F_{Y\to X}^{a}\right]^2\right\}<\frac{3}{m}\left(\frac{m}{n}\right)^2$$

holds in the asymptotic sense, i.e, if $n$ is a sufficiently large. Therefore, in the asymptotic squared meaning, for $m>1$, we have

$$E\left\{\left[F_{Y\to X}^{a}\right]^2\right\}<E\left\{\left[F_{Y\to X}^{c}\right]^2\right\}$$

asymptotically. In other words, the average Granger causality converges to zero more quickly than the cumulative Granger causality.

*Theorem C2. Under the setup as mentioned above,* $\limsup_{n\to\infty} \frac{E\left\{\left[F_{Y\to X}^{a}\right]^2\right\}}{E\left\{\left[F_{Y\to X}^{c}\right]^2\right\}} < 1$ for $m > 1$.

And, the larger $m$ is, the higher asymptotic converge rate the average Granger causality is than the cumulative one.

□

**Appendix D: Dual Kalman filter cumulative Granger causality (Dkf cumulative GC)**

We used the dual Kalman filter as in (Havlicek et al., 2010; Sommerlade et al., 2012), which can be described as the following MVAR

$$\begin{bmatrix} x^{t+1} \\ y^{t+1} \end{bmatrix} = \sum_{k=1}^{p} A_k(t) \begin{bmatrix} x^{t-k} \\ y^{t-k} \end{bmatrix} + \begin{bmatrix} n_1^t \\ n_2^t \end{bmatrix} \quad \text{(D1)}$$

where $A_k(t)$ are the time-varying coefficients and $n_{1,2}^t$ are the white noises. Define

$$z^t = \begin{bmatrix} x^t \\ y^t \end{bmatrix},\ w(t) = \left[(z^t)^T, \cdots, (z^{t-p+1})^T\right]^T,\ a(t) = vec\left(\left[A_1(t)^T, \cdots, A_k(t)^T\right]\right),$$

and Eq. (D1) can be rewritten as

$$z^t = w(t-1)^T a(t) + \eta^t \quad \text{(D2)}$$

associated with a random walk process for the time-varying coefficients

$$a(t+1) = a(t) + \nu^t. \quad \text{(D3)}$$

By the dual Kalman filter approach, the time-varying coefficients can be estimated, and then the residuals of Eq. (D2) can be used to define a cumulative GC by the same fashion as the cumulative GC in the paper

$$dkfGC_{Y\to X} = \log\left(\frac{\sum_{t=1}^{T}\text{var}(\eta_{yx}^{t})}{\sum_{t=1}^{T}\text{var}(\eta_{x}^{t})}\right), \qquad \text{(D4)}$$

where T is the length of the time course, $\eta_{yx}^{t}$ is the noise term in Eq. (D2) for the x-component with considering the inter-dependence from y-component, and $\eta_{x}^{t}$ is the noise term in Eq. (D2) without considering the inter-dependence from y-component. As in (Havlicek et al., 2010), the $dkfGC_{Y\to X}$ and the $dkfGC_{X\to Y}$ can be computed by estimating the model parameters, and the p-value of these causality statistics can be established by bootstrap. The readers are refer to (Havlicek et al., 2010) for more details about the parameter estimation procedure and the bootstrap for significance.

## References

Ahmed, A., Xing, E., 2009. Recovering time-varying networks of dependencies in social and biological studies. Proceedings of the National Academy of Sciences 106, 11878-11883.

Arichi, T., Fagiolo, G., Varela, M., Melendez-Calderon, A., Allievi, A., Merchant, N., Tusor, N., Counsell, S.J., Burdet, E., Beckmann, C.F., Edwards, A.D., 2012. Development of BOLD signal hemodynamic responses in the human brain. NeuroImage 63, 663-673.

Biswal, B., Mennes, M., Zuo, X.-N., et.al., 2010. Toward discovery science of human brain function. Proceedings of the National Academy of Sciences 107, 4734-4739.

Boyacioglu, R., Barth, M., 2012. Generalized iNverse imaging (GIN): Ultrafast fMRI with physiological noise correction. Magn Reson Med Up coming.

Cavanna, A.E., Trimble, M.R., 2006. The precuneus: a review of its functional anatomy and behavioural correlates. Brain. 129, 564-583. Epub 2006 Jan 2006.

Chen, M.-p., Lee, C.-C., Hsu, Y.-C., 2011. The impact of American depositary receipts on the Japanese index: Do industry effect and size effect matter? Economic Modelling 28, 526-539.

Cribben, I., Haraldsdottir, R., Atlas, L.Y., Wager, T.D., Lindquist, M.A., 2012. Dynamic connectivity regression: determining state-related changes in brain connectivity. Neuroimage. 61, 907-920.

Deshpande, G., Sathian, K., Hu, X., 2010. Effect of hemodynamic variability on Granger causality analysis of fMRI. NeuroImage 52, 884-896.

Ding, M., Bressler, S.L., Yang, W., Liang, H., 2000. Short-window spectral analysis of cortical event-related potentials by adaptive multivariate autoregressive modeling: data preprocessing, model validation, and variability assessment. Biological Cybernetics 83, 35-45.

Evan, K., Snigdhansu, C., Auroop R., G., 2011. Exploring Granger causality between global average observed time series of carbon dioxide and temperature. Theoretical and Applied Climatology 104, 325-335.

Feinberg, D.A., Moeller, S., Smith, S.M., Auerbach, E., Ramanna, S., Glasser, M.F., Miller, K.L.,

Ugurbil, K., Yacoub, E., 2010. Multiplexed Echo Planar Imaging for Sub-Second Whole Brain FMRI and Fast Diffusion Imaging. PLoS ONE 5, e15710.

Feinberg, D.A., Yacoub, E., 2012. The rapid development of high speed, resolution and precision in fMRI. NeuroImage 62, 720-725.

Friston, K., Moran, R., Seth, A.K., 2012. Analysing connectivity with Granger causality and dynamic causal modelling. Current Opinion in Neurobiology 23, 1-7.

Friston, K.J., Jezzard, P., Turner, R., 1994. Analysis of functional MRI time-series. Human brain mapping 1, 153-171.

Ge, T., Feng, J., Grabenhorst, F., Rolls, E., 2012. Componential Granger causality, and its application to identifying the source and mechanisms of the top-down biased activation that controls attention to affective vs sensory processing. NeuroImage 59, 1846-1858.

Ge, T., Kendrick, K., Feng, J., 2009. A Novel Extended Granger Causal Model Approach Demonstrates Brain Hemispheric Differences during Face Recognition Learning. PLoS Comput Biol 5, e1000570.

Granger, C.W.J., 1969. Investigating Causal Relations by Econometric Models and Cross-spectral Methods. Econometrica 37, 424-438.

Guo, S., Wu, J., Ding, M., Feng, J., 2008. Uncovering Interactions in the Frequency Domain. PLoS Comput Biol 4, e1000087.

Havlicek, M., Friston, K.J., Jan, J., Brazdil, M., Calhoun, V.D., 2011. Dynamic modeling of neuronal responses in fMRI using cubature Kalman filtering. NeuroImage 56, 2109-2128.

Havlicek, M., Jan, J., Brazdil, M., Calhoun, V.D., 2010. Dynamic Granger causality based on Kalman filter for evaluation of functional network connectivity in fMRI data. Neuroimage. 53, 65-77.

Hemmelmann, D., Ungureanu, M., Hesse, W., Wustenberg, T., Reichenbach, J.R., Witte, O.W., Witte, H., Leistritz, L., 2009. Modelling and analysis of time-variant directed interrelations between brain regions based on BOLD-signals. Neuroimage. 45, 722-737.

Hesse, W., Moller, E., Arnold, M., Schack, B., 2003. The use of time-variant EEG Granger causality for inspecting directed interdependencies of neural assemblies. Journal of Neuroscience Methods 124, 27-44.

Jaynes, E.T., 1985. Some random observations. Synthese 63, 115-138.

Lloyd, S., 1989. Use of mutual information to decrease entropy: implications for the second law of thermodynamics. Physical Review A 39, 5378-5386.

Luo, Q., Ge, T., Feng, J., 2011. Granger causality with signal-dependent noise. NeuroImage 57, 1422-1429.

Luscombe, N.M., Madan Babu, M., Yu, H., Snyder, M., Teichmann, S.A., Gerstein, M., 2004. Genomic analysis of regulatory network dynamics reveals large topological changes. Nature 431, 308-312.

Ohiorhenuan, I.E., Mechler, F., Purpura, K.P., Schmid, A.M., Hu, Q., Victor, J.D., 2010. Sparse coding and high-order correlations in fine-scale cortical networks. Nature 466, 617-621.

Psaradakis, Z., Ravn, M.O., Sola, M., 2005. Markov switching causality and the money–output relationship. Journal of Applied Econometrics 20, 665-683.

Sato, J.o.R., Junior, E.A., Takahashi, D.Y., de Maria Felix, M., Brammer, M.J., Morettin, P.A., 2006. A method to produce evolving functional connectivity maps during the course of an fMRI experiment using wavelet-based time-varying Granger causality. Neuroimage 31, 187-196.

Schippers, M., Renken, R., Keysers, C., 2011. The effect of intra- and inter-subject variability of hemodynamic responses on group level Granger causality analyses. NeuroImage 57, 22-36.

Smerieri, A., Rolls, E.T., Feng, J., 2010. Decision Time, Slow Inhibition, and Theta Rhythm. The

Journal of neuroscience 30, 14173-14181.
Smith, S.M., Bandettini, P.A., Miller, K.L., Behrens, T.E., Friston, K.J., David, O., Liu, T., Woolrich, M.W., Nichols, T.E., 2012. The danger of systematic bias in group-level FMRI-lag-based causality estimation. NeuroImage 59, 1228-1229.
Smith, S.M., Miller, K.L., Salimi-khorshidi, G., Webster, M., Beckmann, C.F., Nichols, T.E., Ramsey, J.D., Woolrich, M.W., 2011. Network modelling methods for FMRI. NeuroImage 54, 875-891.
Sommerlade, L., Thiel, M., Platt, B., Plano, A., Riedel, G., Grebogi, C., Timmer, J., Schelter, B., 2012. Inference of Granger causal time-dependent influences in noisy multivariate time series. J Neurosci Methods 203, 173-185.
Tuncbag, N., Kar, G., Gursoy, A., Keskin, O., Nussinov, R., 2009. Towards inferring time dimensionality in protein-protein interaction networks by integrating structures: the p53 example. Mol. BioSyst. 5, 1770-1778.
Tzourio-Mazoyer, N., Landeau, B., Papathanassiou, D., Crivello, F., Etard, O., Delcroix, N., Mazoyer, B., Joliot, M., 2002. Automated anatomical labeling of activations in SPM using a macroscopic anatomical parcellation of the MNI MRI single-subject brain. Neuroimage. 15, 273-289.
Wang, J., Zhu, H., Fan, J., Giovanello, K., Lin, W., 2011. Adaptively and spatially estimating the hemodynamic response functions in fMRI. Med Image Comput Comput Assist Interv 14, 269-276.
Wen, X., Yao, L., Liu, Y., Ding, M., 2012. Causal interactions in attention networks predict behavioral performance. The Journal of neuroscience 32, 1284-1292.
Wiener, N., 1956. The Theory of Prediction. In: BeckenBach, E. (Ed.), Modern mathematics for the engineer. McGraw-Hill, New York, pp. 165 - 190.
Yan, C., Zang, Y., 2010. DPARSF: a MATLAB toolbox for "pipeline" data analysis of resting-state fMRI. Frontiers in Systems Neuroscience 4.
Zalesky, A., Fornito, A., Harding, I.H., Cocchi, L., Yucel, M., Pantelis, C., Bullmore, E.T., 2010. Whole-brain anatomical networks: does the choice of nodes matter? NeuroImage 50, 970-983.
Zhu, J., Chen, Y., Leonardson, A.S., Wang, K., Lamb, J.R., Emilsson, V., Schadt, E.E., 2010. Characterizing Dynamic Changes in the Human Blood Transcriptional Network. PLoS Comput Biol 6, e1000671.

## Tables and Figure Captions

**Table 1.** The 95% confidence intervals of the differences between the cumulative causality measurements established from time windows with different lengths.

| direction | $X \rightarrow Y$ | | $Y \rightarrow X$ | |
|---|---|---|---|---|
| quantile | 0.025th | 0.975th | 0.025th | 0.975th |
| $D_1^a$ | 0.0078 | 0.0321 | 0.0053 | 0.0341 |
| $D_2^a$ | 0.0085 | 0.0448 | 0.0109 | 0.0453 |
| $D_3^a$ | 0.0002 | 0.0128 | 0.0003 | 0.0229 |
| $D_1^c$ | 0.0078 | 0.0285 | 0.0055 | 0.0333 |
| $D_2^c$ | 0.0105 | 0.0437 | 0.0116 | 0.0480 |
| $D_3^c$ | 0.0002 | 0.0156 | 0.0003 | 0.0264 |

**Table 2. Performance comparison of different methods.** The term 'Classical GC' is short for classic Granger causality, 'Cumulative GC' stands for cumulative Granger causality, 'Average GC' is the average Granger causality, 'Opt Cumulative GC' stands for the cumulative Granger causality with optimally divided time windows, and 'Opt Average GC' indicates the average Granger causality with the optimally determined time windows. $\bar{F}_{X\to Y}$ and $\bar{F}_{Y\to X}$ are the average values of the Granger causality measurements for over 100 simulations, and $\overline{BIC}$ is the mean value of the BIC for 100 simulations. The threshold for significance is $10^{-12}$.

| Method | Window Length | TP | FP | TN | FN | $\bar{F}_{X\to Y}$ | $\bar{F}_{Y\to X}$ | $\overline{BIC}$ |
|---|---|---|---|---|---|---|---|---|
| Classical GC | 1200 | 0% | 0% | 100% | 100% | 0.0023 | 0.0007 | 6.9489 |
| Cumulative GC | 10 | 50% | 0% | 100% | 50% | 0.2484 | 0.1331 | 8.5869 |
| | 50 | 72% | 0% | 100% | 28% | 0.1303 | 0.0208 | 7.2041 |
| | 100 | 75% | 0% | 100% | 25% | 0.1177 | 0.0101 | 7.0188 |
| | 300 | 70% | 0% | 100% | 30% | 0.0709 | 0.0032 | 6.9417 |
| | 600 | 0% | 0% | 100% | 100% | 0.0071 | 0.0015 | 6.9819 |
| Average GC | 10 | 39% | 0% | 100% | 61% | 0.2620 | 0.1516 | 8.5869 |
| | 50 | 85% | 0% | 100% | 15% | 0.1219 | 0.0212 | 7.2041 |
| | 100 | 93% | 0% | 100% | 7% | 0.1094 | 0.0102 | 7.0188 |
| | 300 | 92% | 0% | 100% | 8% | 0.0681 | 0.0032 | 6.9417 |
| | 600 | 14% | 0% | 100% | 86% | 0.0074 | 0.0015 | 6.9819 |
| Opt Cumulative GC | | 90% | 3% | 97% | 10% | 0.1320 | 0.0131 | 6.8780 |
| Opt Average GC | | 94% | 2% | 98% | 6% | 0.1453 | 0.0080 | 6.8780 |

**Table 3. Comparison of running time (in seconds) for different methods**. The classic Granger causality treats the whole time series as one time window. The average Granger causality and cumulative Granger causality were applied to the data by dividing the time series into time windows with equal lengths. The average Granger causality and cumulative Granger causality were also used after optimally dividing the time windows using the proposed algorithm. This simulation was carried out by a computer with an Intel® Core™ 2 CPU T5600 @ 1.83GHz, 1.83GHz, and 1.5G RAM.

| | **One time window** | **Time windows with equal length** | | | | | **Optimally divided time windows** |
|---|---|---|---|---|---|---|---|
| Window length | 1200 | 600 | 300 | 100 | 50 | 10 | |
| Running time | 0.0073 | 0.2936 | 0.4697 | 0.9969 | 1.9747 | 9.8709 | 346.5863 |

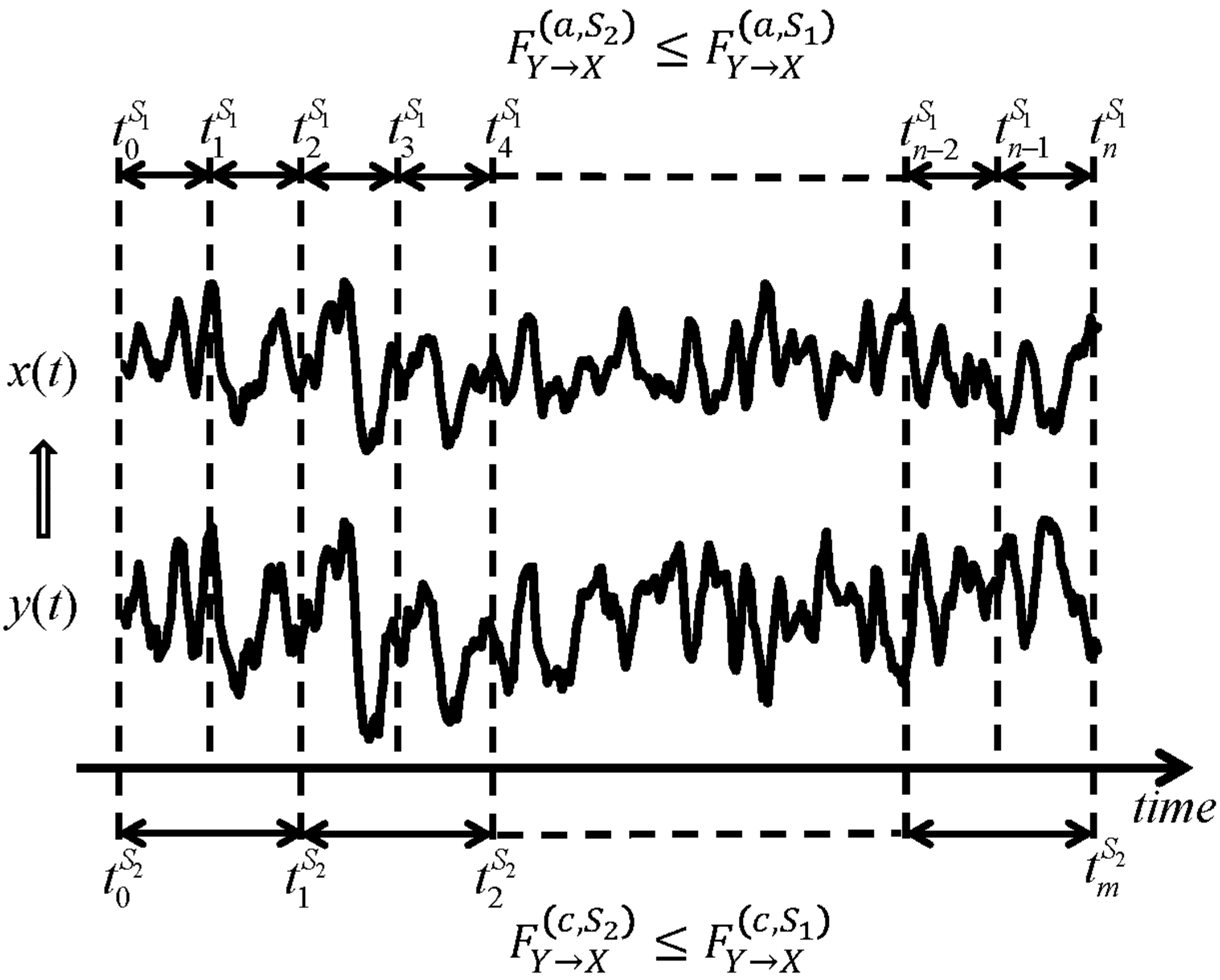

**Figure 1**. Monotonicity of the cumulative and average Granger causalities. If we consider finer time windows with the same length, the change-point set can be derived from the window length, and thus the causality established by different change-point sets can be equivalently denoted by the corresponding window lengths $m_i$ for $S_i$.

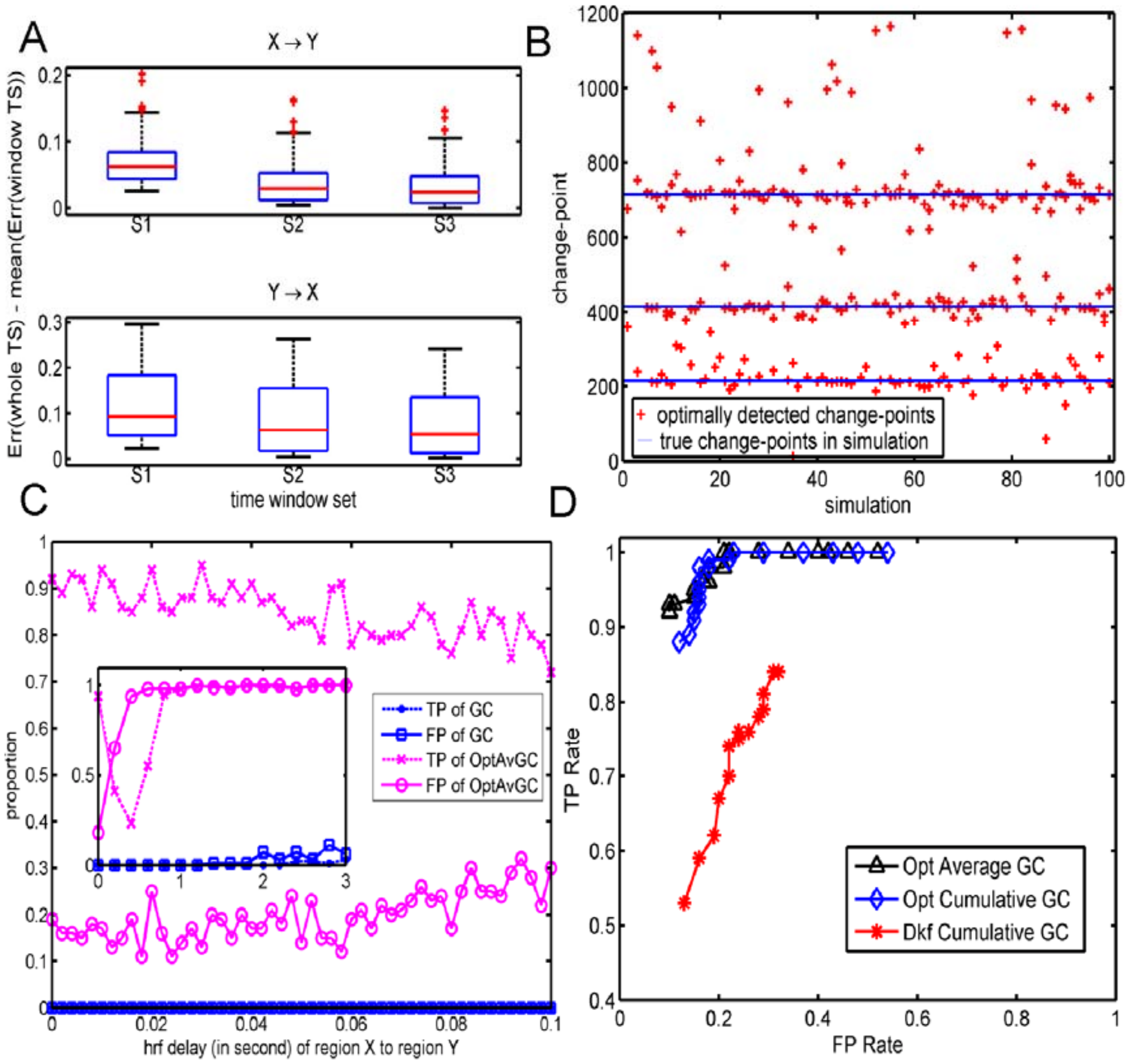

**Figure 2.** Results of the simulation model: (A) Residual variance comparison between the models established from the whole time series and those models fitted on different time window sets. (B) Optimally detected change-points for 100 simulations. (C) Mean of the TP and FP rates given by different methods for each HRF delay in 100 simulations. (D) The TP rate is plotted against the FP rate given by different approaches for each threshold of the p-values in 100 simulations

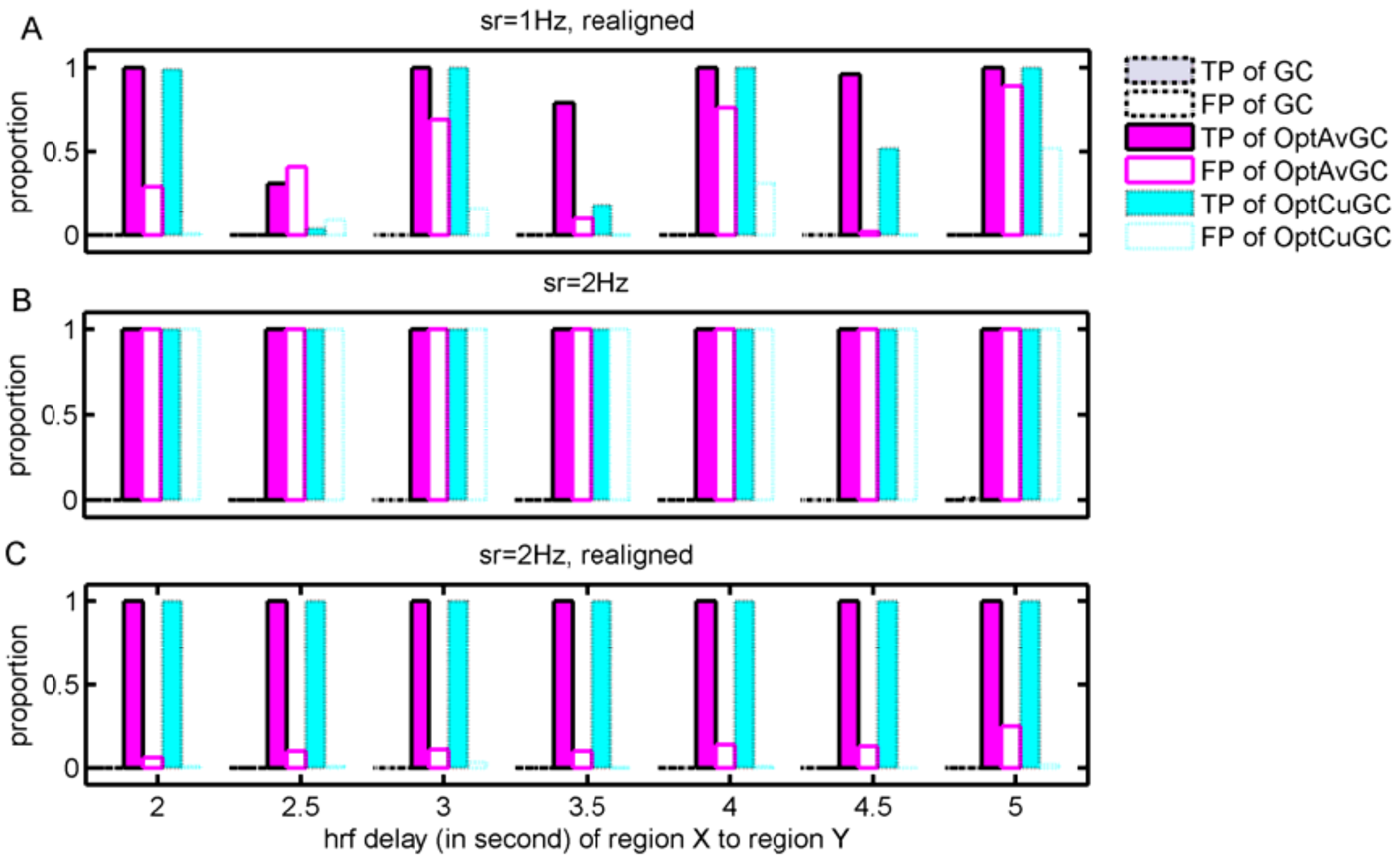

**Figure 3.** Effects of the HRF delay and down sampling on the Granger causality analysis (GCA). (A) Performances of GCAs after a realignment of the BOLD signals between two regions to correct the opposite HRF delay when the sampling rate was 1 Hz. (B) Performances of GCAs when the sampling rate was 2 Hz. (C) Performances of GCAs after a realignment of the BOLD signals when the sampling rate was 2 Hz.

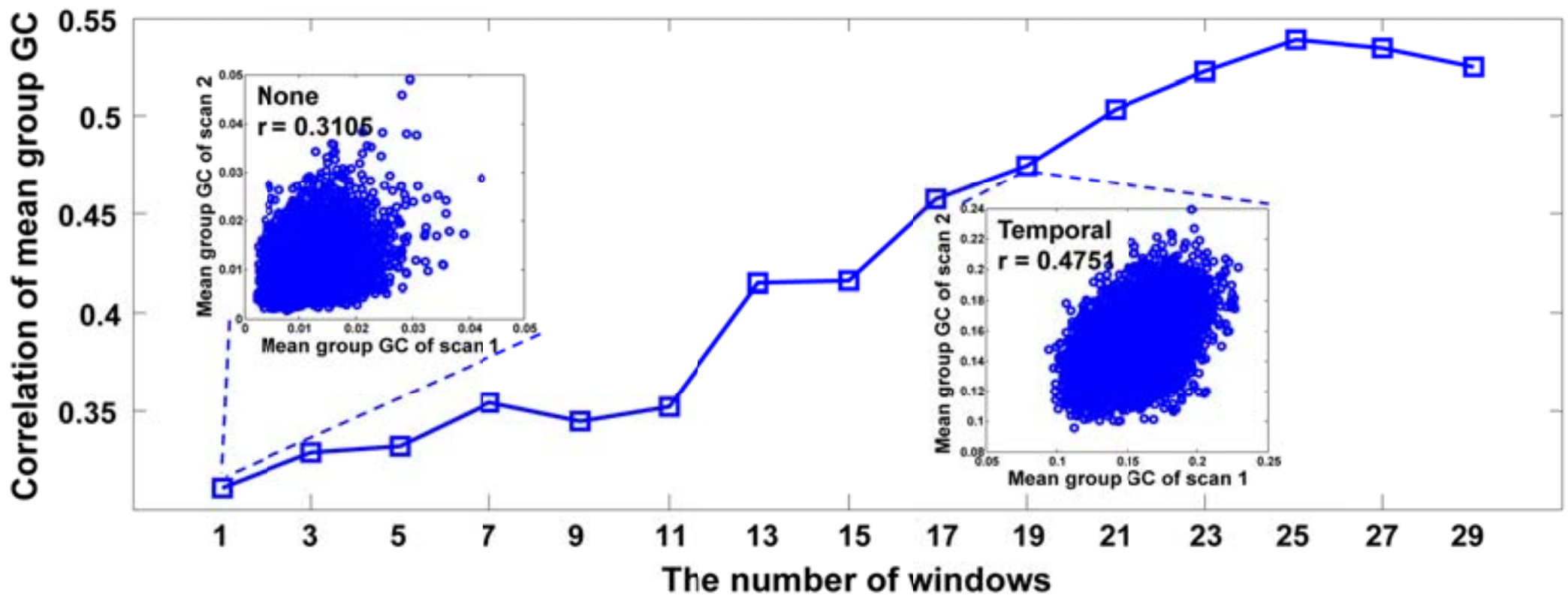

**Figure 4.** Correlation of the mean of the group Granger causalities (GC) between two series of scans upon the same subject set, versus the number of time windows for the average Granger causality. The inset plots show the correlation between two scans of the selected number of windows, where each circle represents the GC between two ROIs parcelled by AAL atlas.

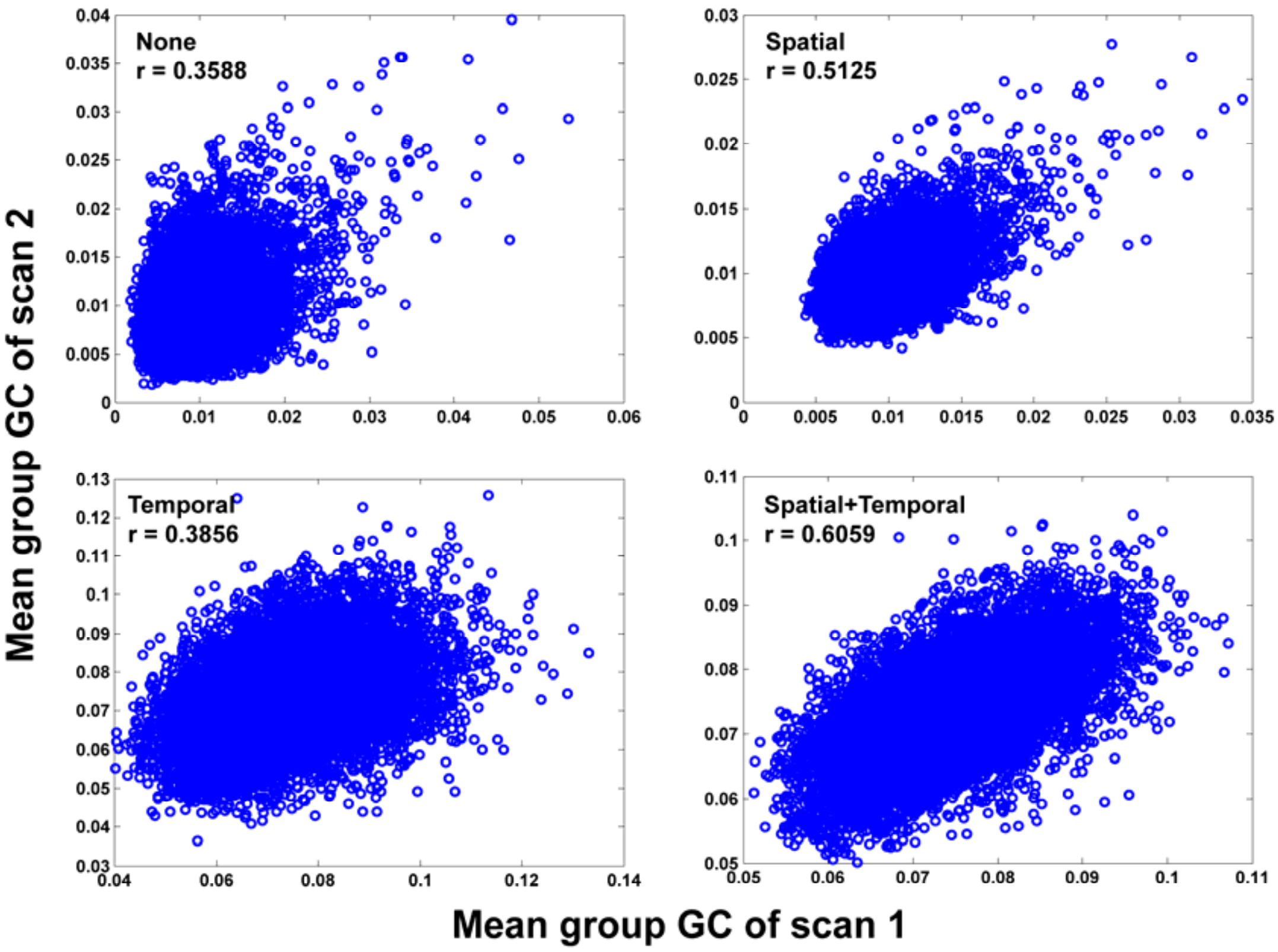

**Figure 5.** Correlation between the Granger causality between two series of scans for the same set of subjects. The causality measurements were calculated through different methods: (A) traditional Granger causality, (B) voxel-level Granger causality, (C) average Granger causality, and (D) spatio-temporal Granger causality.

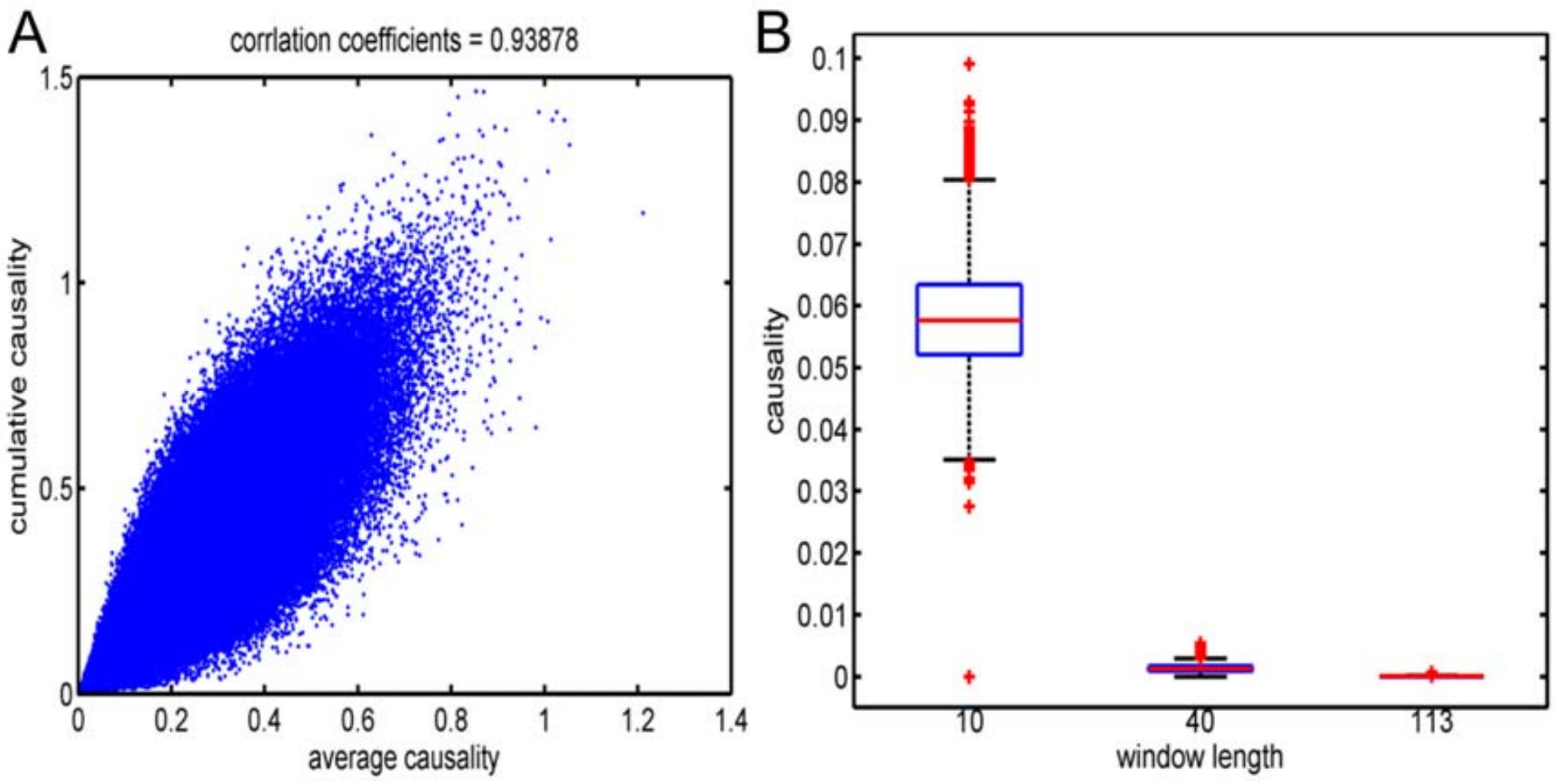

**Figure 6**. Granger causality results of the resting-state dataset. (A) Average Granger

causality versus cumulative Granger causality on the resting state dataset. (B) Comparison of the average Granger causality established by different time window lengths.

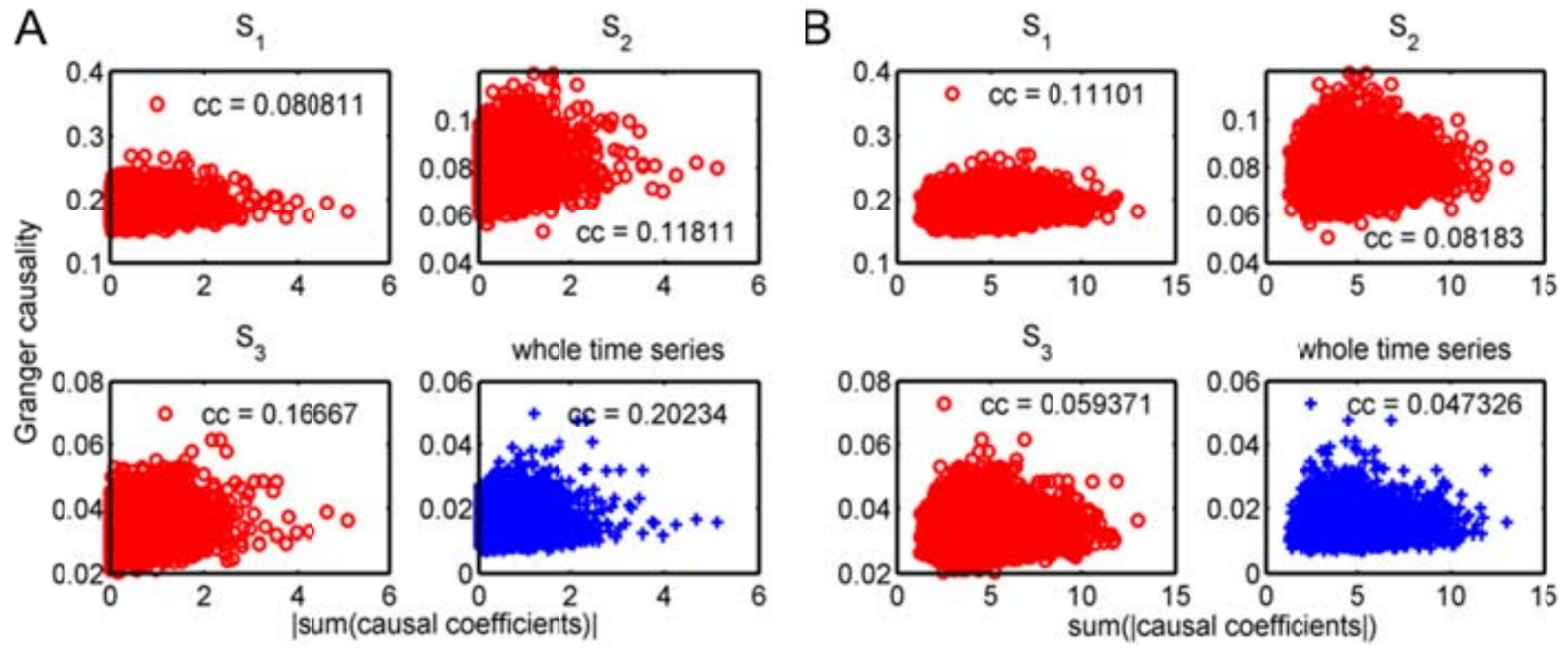

**Figure 7**. Granger causality versus the sum of the causal coefficients across time windows. The causal coefficients were estimated for each time window defined by $S_1$ for each subject. The medians of the causality among 198 subjects were established using different change-point sets, including $S_1, S_2, S_3$, and the whole time series without a change-point (specified as the titles for subplots in the figure). Different change-point sets gave different Granger causality values, since the Granger causality value increased as the time window lengths decreased. (A) Correlation to the absolute value of the median of the sums of the causal coefficients over all time windows. (B) Correlation to the median of the sums of the absolute values of the causal coefficients over all time windows. The *p*-values for all correlations are below the significant threshold.

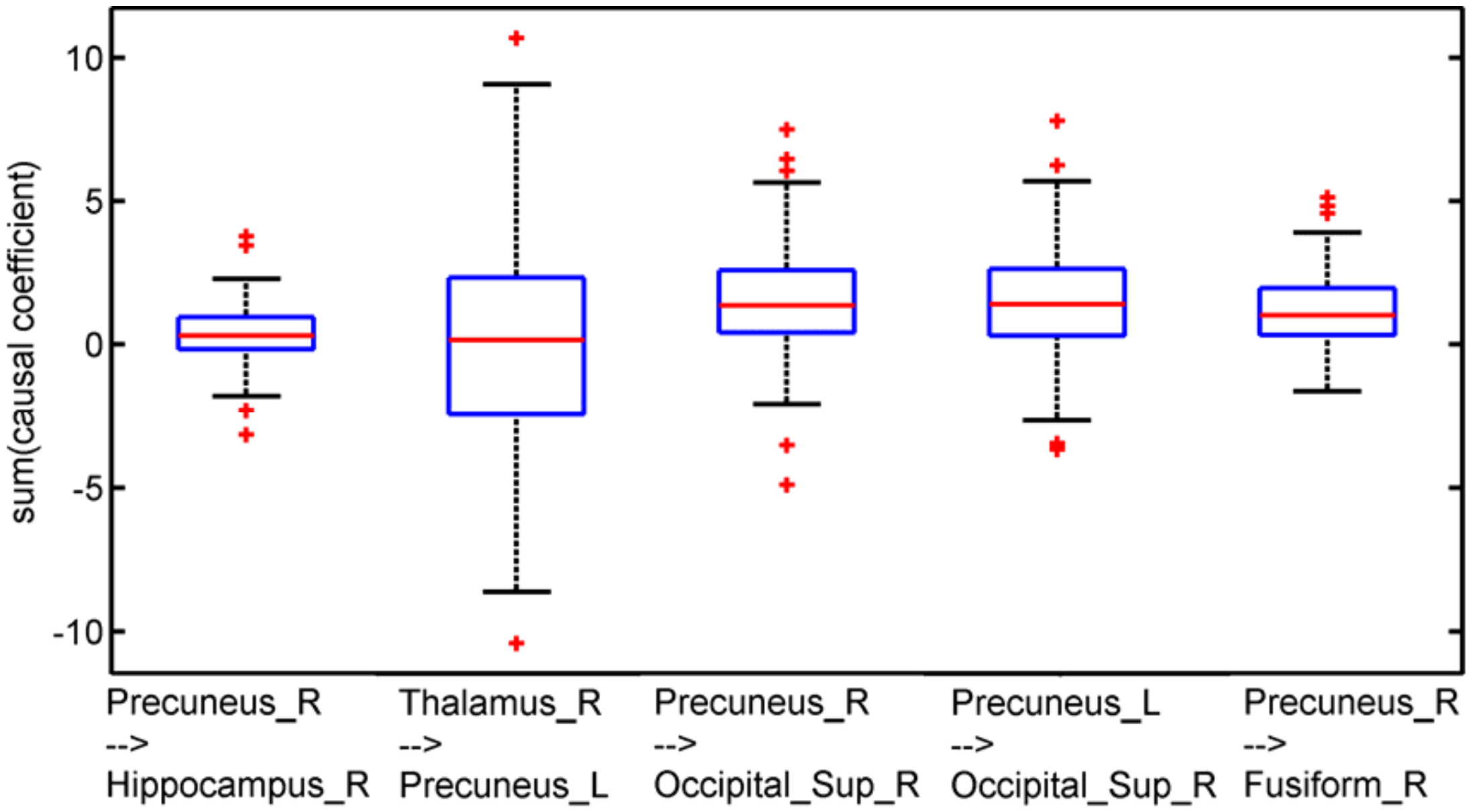

**Figure 8**. Boxplot for the sum of the causal coefficients across all time windows of the change-point set, $S_1$.

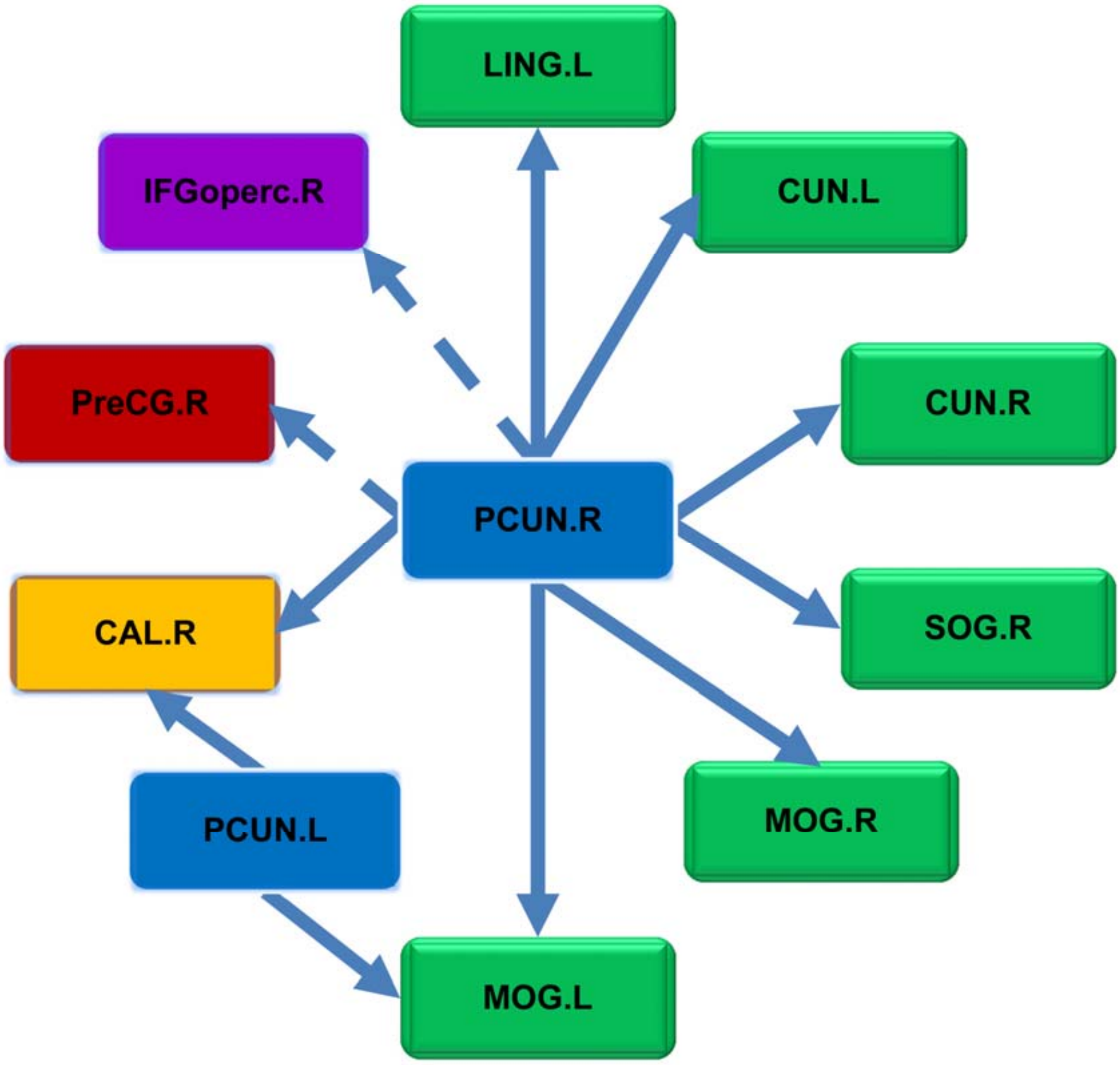

**Figure 9**. Information flows from precuneus inferred by the average Granger causality based on the optimal time window dividing algorithm. The brain regions for visual recognition are marked in green, the primary visual cortex is marked in yellow, the sensory motor areas are marked in red, and the attention areas are marked in purple. The arrows marked by dotted lines indicate potentially false predictions owing to the regional variation of the HRF. The brain regions are defined by AAL90, as in DPARSF (Yan and Zang, 2010).

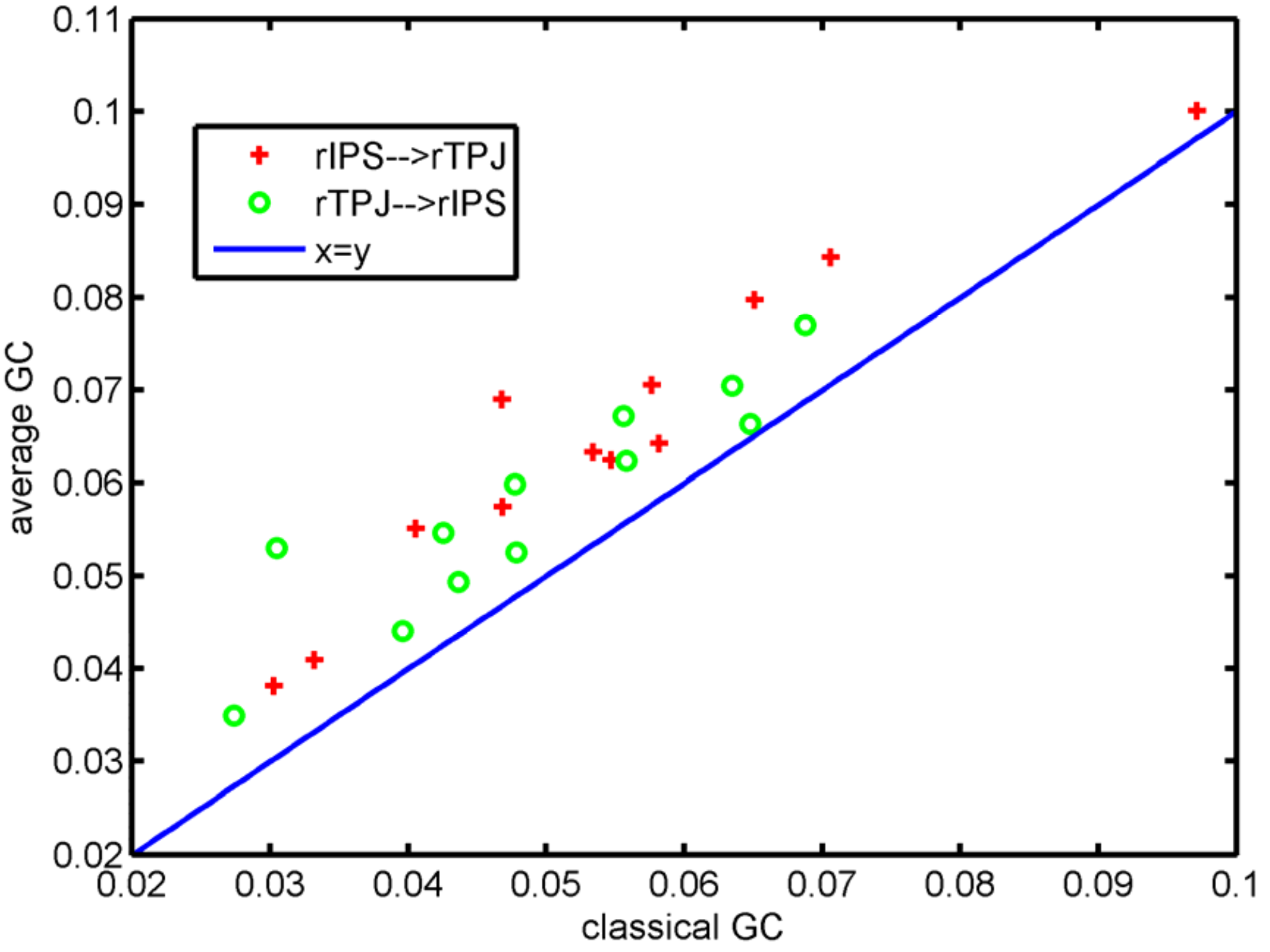

**Figure 10**. Comparison between the classic Granger causality (classical GC) and average Granger causality (average GC) for an attention-task dataset.